\documentclass[11pt]{article}
\usepackage{graphicx}
\usepackage{amssymb}
\usepackage{amscd}
\usepackage{mathrsfs}
\usepackage{longtable,lscape}
\usepackage{amsthm}
\usepackage{amsfonts}
\usepackage{amsmath}
\usepackage{bbm}
\usepackage{float}
\usepackage{url}
\usepackage[round]{natbib}

\newcommand{\compl}{{\mathbb C}}
\newcommand{\real}{{\mathbb R}}
\newcommand{\captionfonts}{\footnotesize}
\makeatletter % Allow the use of @ in command names
\long\def\@makecaption#1#2{%
\vskip\abovecaptionskip
\sbox\@tempboxa{{\captionfonts #1: #2}}%
\ifdim \wd\@tempboxa >\hsize
{\captionfonts #1: #2\par}
\else
\hbox to\hsize{\hfil\box\@tempboxa\hfil}%
\fi
\vskip\belowcaptionskip}
\makeatother 
\begin{document}
\title{The GTR-model: a universal framework for quantum-like measurements}
\author{Diederik Aerts$^1$ and Massimiliano Sassoli de Bianchi$^{2}$ \vspace{0.5 cm} \\ 
\normalsize\itshape
$^1$ Center Leo Apostel for Interdisciplinary Studies \\
\normalsize\itshape
Brussels Free University, 1050 Brussels, Belgium \\ 
\normalsize
E-Mail: \url{diraerts@vub.ac.be}
%\vspace{0.5 cm} 
\\ 
\normalsize\itshape
$^2$ Laboratorio di Autoricerca di Base \\ 
\normalsize\itshape
6914 Lugano, Switzerland \\
\normalsize
E-Mail: \url{autoricerca@gmail.com} \\
}
\date{}
\maketitle
\begin{abstract}
\noindent 
We present a very general geometrico-dynamical description of physical or more abstract entities, called the \emph{general tension-reduction} (GTR) model, where not only states, but also measurement-interactions can be represented, and the associated outcome probabilities calculated. Underlying the model is the hypothesis that indeterminism manifests as a consequence of unavoidable fluctuations in the experimental context, in accordance with the \emph{hidden-measurements interpretation} of quantum mechanics. When the structure of the state space is Hilbertian, and measurements are of the \emph{universal} kind, i.e., are the result of an average over all possible ways of selecting an outcome, the GTR-model provides the same predictions of the Born rule, and therefore provides a natural completed version of quantum mechanics. However, when the structure of the state space is non-Hilbertian and/or not all possible ways of selecting an outcome are available to be actualized, the predictions of the model generally differ from the quantum ones, especially when sequential measurements are considered. Some paradigmatic examples will be discussed, taken from physics and human cognition. Particular attention will be given to some known psychological effects, like question order effects and response replicability, which we show are able to generate non-Hilbertian statistics. We also suggest a realistic interpretation of the GTR-model, when applied to human cognition and decision, which we think could become the generally adopted interpretative framework in quantum cognition research.
\end{abstract}
\\

\section{Introduction}

Probability is the key notion used by scientists of different disciplines to quantify, in a meaningful and optimal way, their lack of knowledge regarding certain properties of the systems under study. Before the advent of quantum mechanics, only classical probabilities were taken into consideration, based on the structure of Boolean algebra and obeying the classical Kolmogorovian axioms~\citep{Kolmogoroff1933}. On the other hand, quantum probabilities, which are based on a different structure of the experimental propositions, cannot be represented in a fixed probability space (if multiple measurements are considered), and therefore generally disobey the Kolmogorovian axioms. 

However, the difference between classical and quantum probabilities is not in a simple correspondence with the difference between macroscopic and microscopic physical entities. Indeed, also macroscopic entities can behave in a quantum-like way, depending on the nature of the experimental actions that an experimenter is considering in relation to them. In other terms, quantum and classical probabilities can always emerge from our experimental investigations, depending on the \emph{structure of possibilities} that are taken into consideration. 

This means that classic and quantum probabilities should be considered as special cases of more general probability structures, which can simply be called (depending on the context) non-classical, non-quantum, or quantum-like. These more general structures are non-classical in the sense that they do not obey the Kolmogorovian axioms, and they are quantum-like in the sense that, similarly to quantum probabilities, they are based on logical connectives that are \emph{dynamical}, i.e., describing the possible outcomes of \emph{actions} (measurements) that can be performed on the different entities. However, they are also non-quantum, in the sense that they are not necessarily purely quantum, as the structure of the associated state space is not necessarily Hilbertian and the probability values are not necessarily those predicted by the Born rule. 

From the viewpoint of physics, the interest of investigating more general probability models lies for instance in the possibility of shedding new light onto the problem of the semiclassical limit, i.e., in the understanding of the transition from pure quantum to pure classical regimes. Indeed, if it is true that classical and quantum probabilities are based on different, not commensurable structures, obtained as different limits of more general quantum-like situations, then it is also clear that we need more general models, of a ``mixed'' quantum-classical kind, if we want to describe the mesoscopic regions of our reality, which cannot be incorporated within the pure quantum or pure classical limit models. 

A preliminary analysis of these intermediary regions of reality, described by non-classic and non-quantum probabilities, was carried out in some detail by one of us and his collaborators in the past decades, using a paradigmatic model called the $\epsilon$-model \citep{Aerts1998, SassolideBianchi2013a}: a generalization of a two-level (qubit) system where an additional real parameter $\epsilon$ can be continuously varied, from $0$ to $1$, so as to produce a non-singular classic-to-quantum transition. Even though the initial motivation in studying these more general probability models came from physics, it became clear early on that their significance went beyond the field of physics, and was of great interest also in the description of human cognitive processes, as today studied in the new emerging field of theoretical and experimental investigation called \emph{quantum cognition} \citep{Kitto2008, Khrennikov2010,BusemeyerBruza2012, AertsEtal2013a, AertsEtal2013b, PothosBusemeyer2013, WangEtal2013, BlutnerbeimGraben2014}.

Quantum cognition resulted, among other things, from the observation that human concepts, understood as abstract entities interacting with human minds that are sensitive to their meanings, can produce highly contextual dynamics, impossible to explain by only using traditional modelizations in terms of logico-rational thinking processes, and therefore classical probabilities. Similarly, quantum mechanics was historically created, as a mathematical theory, to offer a consistent description of entities whose behavior appeared also to be highly contextual. Therefore, it was quite natural at some point to assume that the quantum formalism could also play a role in the modelization of human cognition and decision-making. 

This intuition was followed by an increasing number of successful applications of the quantum formalism, and today it is a well-established hypothesis that, in addition to our classical logical layer, describable by classical probability theory, there is an additional quantum conceptual layer, describable by standard quantum mechanics \citep{AertsSozzoVeloz2015}. This quantum description of the human cognitive behavior, however, has nothing to do with the fact that our human brains would be quantum machines. Quantum cognition is not concerned with the modeling of  human brains as quantum computers, but with the possibility of using  quantum probabilities, and the structure of Hilbert spaces, to elucidate the working of our mental processes, particularly those mistakenly understood as irrational.

Clearly, the contextuality of human concepts mirrors that of elementary quantum entities. Consider for instance the problem of concept combination, i.e., the problem that not all concept combinations will have an intersective semantics, so that the meaning of a combination of concepts will not always be reducible to the meaning of the individual concepts forming the combination, as new meanings are constantly able to emerge. This emergence effect can be naturally described within the quantum formalism by means of the superposition principle and the related constructive and destructive interference effects, which can explain the observed overextension and underextension of the probabilities (with respect to the classical predictions). The superposition principle is in turn a key ingredient in the creation of entangled states that can be used to describe the situation of concepts connected through meaning, which are  able to violate Bell's inequalities in a way similar to quantum microscopic entities \citep{Aertsetal2000,AertsSozzo2011}. Other fundamental aspects of the quantum formalism, like for instance quantum field many particles dynamics, can be also be exploited to describe typical situations where human judgments and decisions are at play \citep{AertsSozzo2015}.

This ``unreasonable'' success of quantum mathematics in the modelization of human cognitive and decision situations requires of course to be widely explained. At the same time, one needs to investigate what are its limits, i.e., to what extent the standard quantum formalism can be used to model all sorts of cognitive situations. These two issues are intimately related. Indeed, there are no a priori reasons for the contextuality built-in in the standard Hilbertian formalism to be exactly the same (in terms of structure) as that incorporated in our human mental processes, also because the latters describe a (non-physical, mental) layer of our reality which, evolutionarily speaking, is much younger than the layer of the fundamental physical processes. This means that, starting from a more general model, containing both the quantum and classical regimes as special situations, one should be able to explain why certain aspects of the quantum formalism, in particular the Born rule, are so effective in describing many empirical data, and at the same time insufficient to model many others, considering that a Hilbert space, equipped with the Born rule, necessarily imposes some specific constraints (like the QQ-equality introduced by \citet{WangBusemeyer2013}), that can be violated by our complex cognitive and decisional processes. 

To provide a convincing explanation of both the success of quantum probabilities, and their lack of universality, the present article is organized as follows. In Sec.~\ref{Measuring a coin}, we use the simple example of a coin flipping experiment to motivate a general description of an entity that gives rise to a general measurement model, called the \emph{general tension-reduction model} (GTR-model). The model was recently derived in the ambit of quantum cognition studies \citep{AertsSassolideBianchi2015a,AertsSassolideBianchi2015b}), but is of great interest also for physics, as it offers a non-circular derivation of the Born rule and therefore constitutes a possible solution to the measurement problem \citep{AertsSassolideBianchi2014a,AertsSassolideBianchi2015c}. 

Our strategy here is not that of repeating our previous more formal derivations of the model, but to motivate its construction starting from more qualitative and general considerations. More precisely, inspired by our analysis of the coin flipping measurement of Sec.~\ref{Measuring a coin}, the GTR-model will be introduced in Sec.~\ref{The general tension-reduction (GTR) model}. In Sec.~\ref{Degenerate measurements in the GTR-model}, it will be shown to  allow for the description of experiments where some of the outcomes can be degenerate and, in Sec.~\ref{Composite entities in the GTR-model}, we show that it can naturally handle also the situation of composite entities. In Sec.~\ref{The quantum mechanical example}, we explain how the Born rule can be deduced, when a huge (universal) average is performed over all possible kinds of measurements, showing that the Born rule can be interpreted as a first order approximation of a more general theory, thus explaining its great success also in the description of cognitive experiments. In Sec.~\ref{Non-Kolmogorovian non-Hilbertian structures}, we explicitly show that the GTR-model can describe more general structures than the Kolmogorovian and Hilbertian ones and, in Sec.~\ref{The human cognition example}, we apply the model to human cognition, showing that one needs its full structural richness to described some of the experimental data, like question order effects and response replicability. Finally, in Sec.~\ref{Conclusion}, we offer a few concluding remarks.

\section{Measuring a coin}
\label{Measuring a coin}

Consider the process that consists in flipping a coin onto the floor. If we are interested in knowing the final upper face of the coin, three possible outcomes have to be distinguished: ``head,'' ``tail'' and ``edge,'' and to these three outcomes three different probabilities can be associated: $P({\rm h})$, $P({\rm t})$ and $P({\rm e})$. In the case of an \emph{American nickel}, their typical experimental values are \citep{Murray1993}: 
\begin{equation}
P({\rm h})={2999.5\over 6000},\quad P({\rm t})={2999.5\over 6000}, \quad P({\rm e})={1\over 6000}.
\label{American nickel}
\end{equation} 

Flipping a coin onto the floor is a simple action producing a non-predeterminable outcome, and the same is true when we toss a die, draw a ball from a urn, etc. All these simple experiments, called \emph{chance games}, have been largely used in the past to study the logic of probabilities, and culminated in modern classical probability theory, axiomatized by \citet{Kolmogoroff1933}. However, the actions associated with chance games are of a very special kind, and one should expect a probability model derived from their analysis to also be a very special model, not necessarily able to describe all the probability structures that can emerge from our experiments.

To explain in which sense the random processes traditionally studied by classical probability theory are special, take the coin flipping example. The associated probabilities will depend in part on how the coin is manufactured. For instance, if we have exactly three possible outcomes, this is because there are three distinct faces (two flat faces and a curved one), and the values of the probabilities certainly also depend, in part, on their relative surfaces, on the exact location of the center of mass of the coin, and so on. These are the so-called \emph{intrinsic} properties of the coin, i.e., the attributes it always possesses, in a stable and permanent way (at least for as long as the coin exists). 

But a coin, as a physical entity, is not only described by its intrinsic (always actual) properties: its condition is also determined by those properties that can contextually change over time, like its position and orientation in space, its linear and angular momentum, its temperature, etc. Now, a process like that of flipping a coin is special because it is usually so conceived that we cannot learn anything about its non-intrinsic properties from the obtained probabilities. In other terms, flipping a coin is an experiment which tells us nothing about the \emph{state} of the coin prior to its execution: there is no \emph{discovery aspect} involved, but only a \emph{creation aspect}. 

In \citep{AertsSassolideBianchi2015a,AertsSassolideBianchi2015b}, we have named these special -- creation only -- processes, \emph{solipsistic measurements}, with the term ``solipsistic'' used in a metaphorical sense, to express the idea that the measurement tells us nothing about the pre-measurement state of the entity, but only about the measurement process itself. Note that in the following we shall use the terms ``measurement'' and ``experiment'' almost interchangeably, being clear that a measurement is an experiment aimed at the observation of a given quantity, and that also the flipping of a coin can be interpreted as a measurement process, whose outcomes are the values taken by a quantity called the ``upper face'' of the coin. 

It is worth observing that solipsistic measurements have a truly remarkable property: being totally insensitive to the initial state of the entity,\footnote{It is because solipsistic measurements are processes that are extremely sensitive to small fluctuations that, statistically speaking, they are totally insensitive to variations in the initial state of the entity.} they are necessarily all \emph{mutually compatible experiments}. Not in the sense that they can be performed at the same time (almost no experiments have this remarkable property), but in the sense that the order with which they are carried out is irrelevant. This because the final state obtained by the first measurement cannot influence the outcome of the second one, and vice versa. It is then clear why the Kolmogorovian probability model, which is founded on the paradigm of the solipsistic measurements, is unable to account for the quantum probabilities: quantum measurements, from which quantum probabilities result, are non-solipsistic indeterministic processes, producing statistics of outcomes which strongly depend on the initial pre-measurement states, and therefore cannot in general be mutually compatible measurements.

Let us exploit a bit further the evocative example of the coin to see how we can go from solipsistic measurements to a more general class of measurements, to model more general (non-Kolmogorovian) probability situations. Clearly, an experimenter is free to conceive different measurements, by simply defining different observational protocols. In the case of the coin, one can for instance consider different ways to produce its flipping. Solipsistic measurements, as is known, correspond to a situation where the experimenter has to flip the coin with a vigorous momentum, on a sufficiently hard floor. On the other hand, if the experimenter decides to flip the coin in a less vigorous way, or onto a softer surface, or even a sticky one, it is easy to imagine that the statistics of outcomes will start depending on how the coin is initially positioned before the flipping, for instance on the bottom of the dice cup that is used to produce the shot. 

To model this possibility, we need to find a representation that allows us to express a dependency of the probabilities on the pre-measurement state. A very simple idea would be to represent the state of the coin directly in terms of the associated outcome probabilities. Of course, by doing so we will not be able anymore to describe solipsistic measurements (as is clear that for them the different initial states are all associated with the same outcome probabilities), but let us explore anyway this idea, as it will show us the path for its natural generalization.

So, let us assume that we have chosen a given flipping protocol (i.e., a given measurement), and that by repeating the experiment many times, with the coin always in the same initial state inside the cup, we have obtained the three outcome probabilities $P({\rm h})$, $P({\rm t})$ and $P({\rm e})$. The idea is to describe the initial state of the coin as an abstract point-particle in $\real^3$, with position ${\bf x}=(P({\rm h}), P({\rm t}), P({\rm e}))$, i.e., as a point-particle whose coordinates are precisely the outcome probabilities. Since $P({\rm h})+P({\rm t})+P({\rm e})=1$, it follows that ${\bf x}$ belongs to a two-dimensional regular simplex $\triangle_2$, i.e., to an equilateral triangle of side $\sqrt{2}$. 

We obtain in this way a simple and natural geometric representation of the probabilities characterizing the measurement under investigation. However, as we said, this representation cannot be used to describe solipsistic measurements, as two states ${\bf x}$ and ${\bf x}'$, if different, will necessarily be associated with different outcome probabilities. In fact, this representation doesn't allow, neither, to describe \emph{deterministic} measurements, such that for a given initial state the outcome would be predetermined. These are experiments such that the protocol allow the experimenter to flip the coin in a perfectly controlled (fluctuation free) way, so as to know in advance what will be the final state, given the initial one.

\begin{figure}[!ht]
\centering
\includegraphics[scale =.55]{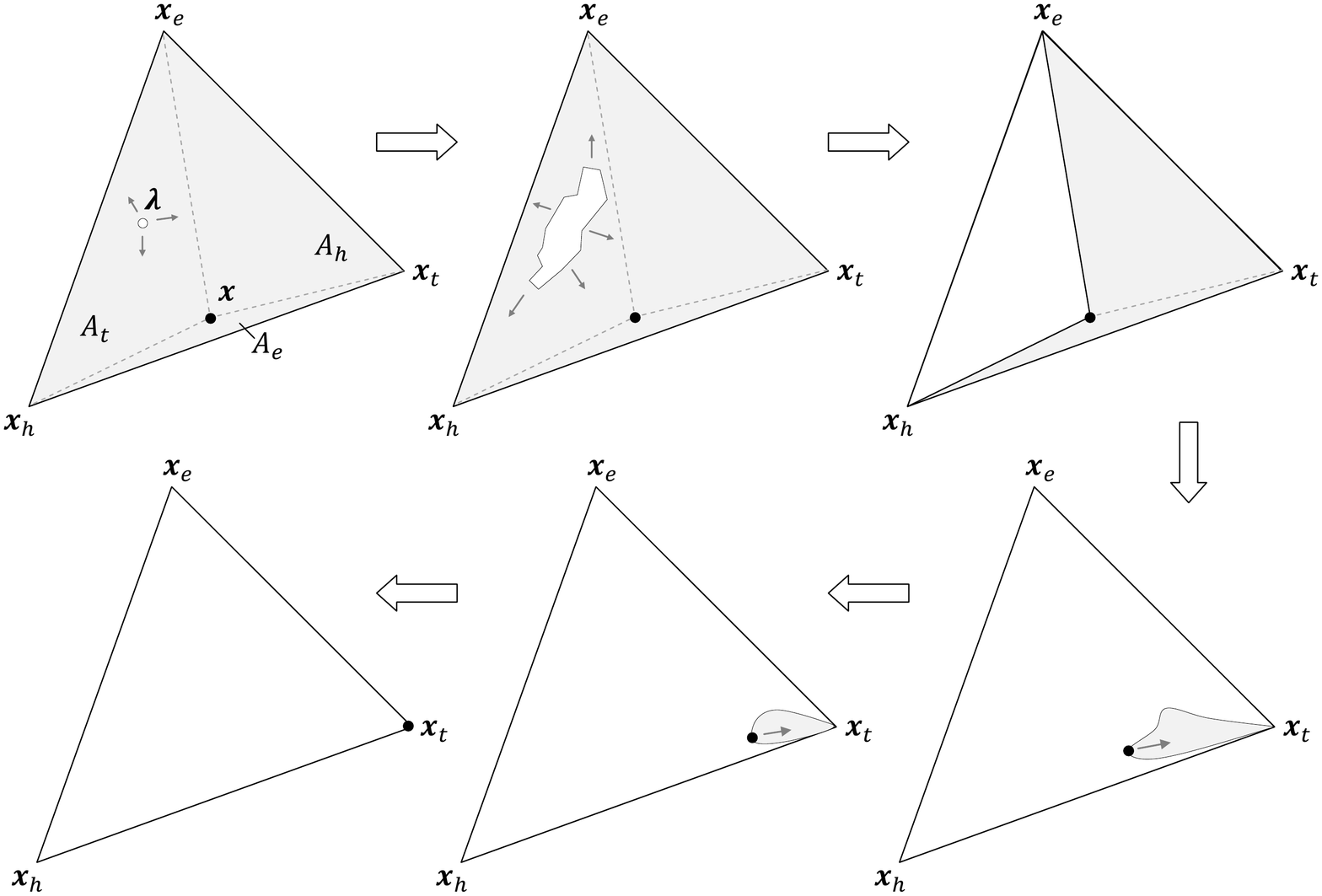}
\caption{A $2$-dimensional triangular membrane stretched between the three vertex points ${\bf{x}}_{\rm h}$, ${\bf{x}}_{\rm t}$ and ${\bf{x}}_{\rm e}$, with the point particle attached to it at point ${\bf x}$, giving rise to three disjoint convex regions $A_{\rm h}$, $A_{\rm t}$, and $A_{\rm e}$. The vector \mbox{\boldmath$\lambda$}, here assumed to belong to region $A_{\rm t}$, indicates the initial point of disintegration of the membrane, which by collapsing brings the point particle to point ${\bf{x}}_{\rm t}$, corresponding to the final outcome of the measurement.
\label{triangolo}}
\end{figure}

If we want to obtain a representation that can be used to also represent solipsistic and deterministic measurements, we thus need to find a way to describe the state of the entity independently from the outcome probabilities. For this, we need to introduce in our model some additional \emph{elements of reality}, describing the \emph{interactions} between the measured entity and the measuring system, i.e., between the coin and the cup-floor system, in our example. To do so, we start by observing that the point ${\bf x}$ exactly defines three disjoint triangular sub-regions in $\triangle_2$, which we will call $A_{\rm h}$, $A_{\rm t}$ and $A_{\rm e}$ (see the first drawing of Fig.~\ref{triangolo}). A simple geometric calculation then shows that \citep{Aerts1986,AertsSassolideBianchi2014a,AertsSassolideBianchi2015a,AertsSassolideBianchi2015b}:
\begin{equation}
P({\rm h})={\mu(A_{\rm h})\over \mu(\triangle_2)}, \quad P({\rm t})={\mu(A_{\rm t})\over \mu(\triangle_2)},\quad P({\rm e})={\mu(A_{\rm e})\over \mu(\triangle_2)},
\label{relative Lebesgue}
\end{equation} 
where $\mu$ denotes the Lebesgue measure. In other terms, the relative areas of the three sub-regions defined by the state-vector ${\bf x}$ are exactly the outcome probabilities. To exploit this remarkable geometric property of simplexes, we are now going to describe a \emph{tension-reduction process} that will allow us to represent, in an abstract way, the (possibly) indeterministic part of a measurement process. 

Assume that $\triangle_2$ is an elastic and disintegrable membrane, stretched between its three vertex points -- let us call them ${\bf x}_{\rm h}$, ${\bf x}_{\rm t}$ and ${\bf x}_{\rm e}$, respectively -- and that the state of the entity is represented by a point-particle firmly attached to the membrane, at some point ${\bf x}$. Assume also that the line segments separating the three regions $A_{\rm h}$, $A_{\rm t}$ and $A_{\rm e}$ are like ``tension lines," along which the membrane can less easily disintegrate. Then, consider the following process: the membrane disintegrates, at some unpredictable point \mbox{\boldmath$\lambda$}. If $\mbox{\boldmath$\lambda$}\in A_{\rm h}$, the disintegrative process propagates inside the entire sub-region $A_{\rm h}$, but not in the other two sub-regions $A_{\rm t}$ and $A_{\rm e}$, because of the tension lines. This will cause the two anchor points of $A_{\rm h}$, ${\bf x}_{\rm t}$ and ${\bf x}_{\rm e}$, to tear away, and being the membrane elastic, it will consequently collapse toward the remaining anchor point ${\bf x}_{\rm h}$, drawing in this way the point particle (which is attached to it) to the same final position, representative of the outcome ``head.''

Similarly, if the initial disintegration point happens in $A_{\rm t}$, the final outcome will be ``tail,'' represented by the vector ${\bf x}_{\rm t}$ (see Fig.~\ref{triangolo}), and if the initial disintegration point happens in $A_{\rm e}$, the final outcome will be ``edge,'' represented by the vector ${\bf x}_{\rm e}$. If the membrane is uniform, it is clear that the probability to obtain outcome ${\bf x}_{\rm h}$ is just given by the relative area of sub-region $A_{\rm h}$, which according to (\ref{relative Lebesgue}) corresponds to the first component of state vector ${\bf x}$, and similarly for the other two outcomes.

Of course, a uniformly disintegrable membrane is a very special situation, and a priori an (uncountable) infinity of different membranes, characterized by different \emph{ways of disintegrating}, can be considered, like for instance those giving rise to solipsistic measurements, whose outcome probabilities are independent of the initial state ${\bf x}$. These can be understood as the limit of membranes that become less and less disintegrable in their interior points and more and more disintegrable in the points belonging only to their three edges, so that the position of the point particle on $\triangle_2$, representative of the initial state, becomes irrelevant in the determination of the outcome probabilities, which will only depend on the probabilities that the initial disintegration happens in one of the three edges of $\triangle_2$. More precisely, if the disintegration probability of the edge opposite to ${\bf x}_{\rm h}$ is $P({\rm h})$, then this will also be the probability for outcome ${\bf x}_{\rm t}$, for (almost) all initial states, and similarly for the other two edges. 

The membranes describing deterministic measurements, on the other hand, can be understood as the limit of membranes becoming more and more disintegrable in a sub-region that becomes increasingly small and less and less disintegrable everywhere else. Indeed, in this limit we obtain a structure which almost surely will start disintegrating in a single predetermined point \mbox{\boldmath$\lambda$}. Then, for almost all initial states one can predict the outcome in advance, with certainty. Indeed, if the initial state ${\bf x}$ is such that $\mbox{\boldmath$\lambda$}\in A_{\rm h}$, the outcome will be ${\bf x}_{\rm h}$, with probability $P({\rm h})=1$, and similarly for the other two outcomes. Note that if we have said that outcomes are predetermined for \emph{almost} all, and not all initial states, this is because we cannot exclude the special situation ${\bf x}= \mbox{\boldmath$\lambda$}$, of classical unstable equilibrium, which remains clearly indeterminate (but do not contribute to the probability calculus, being this possibility of zero Lebesgue measure).

To describe the most general typology of disintegrable membrane, we only need to introduce a \emph{probability density} $\rho:\triangle_2\to [0,\infty[$, $\int_{\triangle_2}\rho({\bf y})d{\bf y} =1$, characterizing the propensity of the membrane (which will be called $\rho$\emph{-membrane}) to disintegrate in its different possible sub-regions. This means that if the initial state produces the three sub-regions $A_{\rm h}$, $A_{\rm t}$ and $A_{\rm e}$, the probabilities for obtaining the three outcomes ${\bf x}_{\rm h}$, ${\bf x}_{\rm t}$ and ${\bf x}_{\rm e}$, will be given by the integrals: 
\begin{equation}
P({\bf x}\to {\bf x}_i|\rho) = \int_{A_i} \rho({\bf{y}})d{\bf{y}}, \quad i\in \{{\rm h},{\rm t}, {\rm e}\},
\label{three integrals}
\end{equation} 
and of course, in the special case of a uniform membrane, we simply have $\rho({\bf{y}}) ={1\over \mu(\triangle_2)}={2\over \sqrt{3}}$, for all ${\bf{y}}$, and we recover~(\ref{relative Lebesgue}).

Before continuing in the construction of our model, a remark is in order. It is clear that the `tension-reduction' mechanism associated with the disintegrable elastic membranes can only describe idealized measurements of the \emph{first kind}, i.e., measurements such that, if repeated a second time, will produce exactly the same outcome, with probability 1. This is so because if ${\bf x}$ corresponds to one of the three vertices of $\triangle_2$, then, being already located in one of the end points of the elastic structure, its position cannot be altered by a new membrane's collapse, when the measurement process is repeated. 

Of course, not all measurements are of the first kind, but certainly most of them can be made, at least ideally, of the first kind. For instance, to make the coin flipping a measurement of the first kind it is sufficient to specify in the protocol that if the coin is already located on the floor, i.e., if the initial state of the coin is an \emph{on-floor state}, and not an \emph{on-cup state}, then what the experimenter has to do is to simply observe if the upper face is ``head,'' ``tail'' or ``edge,'' and take the result of such observation as the outcome of the measurement.

Clearly, the flipping of the coin producing the three on-floor outcomes ``head,'' ``tail'' and ``edge,'' is not the only coin-measurement that we can perform. Imagine for a moment the following \emph{coin shaking} measurement, operationally defined by the following protocol: If the coin is on the floor, then put it at the center of the bottom of the cup, with exactly the same upper face, then shake the cup, following a predetermined procedure (that we don't need to specify here), and finally look at the bottom of the cup, to see what is the obtained upper face. On the other hand, if the coin is already in the cup, just observe its upper face, which will then be the outcome of the measurement. 

We now have two different measurements, the \emph{coin-flipping} measurement, which can produce the outcome states ``floor-head,'' ``floor-tail'' and ``floor-edge,'' and the \emph{coin-shaking} measurement, which can produce the outcome states ``cup-head,'' ``cup-tail'' and ``cup-edge,'' and of course we cannot associate the same membrane (i.e., the same measurement simplex) to both measurements. This not only because their outcome states are different, but also because the associated flipping and shaking procedures are different.

If different membranes can represent different measurements, belonging to a same state space, then in addition to the tension-reduction process describing the membrane's collapse we have to introduce a mechanism allowing the states belonging to one membrane to be measured with respect to another membrane. In other terms, we need to describe a process that can transform an off-membrane state into an on-membrane state, in order to be subjected to its (possibly) indeterministic collapse. A process of this kind has to be able, in particular, to bring the state ``cup-head'' in contact with the `potentiality region' of the flipping-membrane, or the state ``floor-head'' in contact with the `potentiality region' of the shaking-membrane. 

Since we are here interested in obtaining a geometrical representation, and that we want the description to be as simple as possible, a very natural choice is to use a (deterministic) \emph{orthogonal projection} process. In other terms, if ${\bf x}$ is an off-membrane state, with respect to the considered measurement, we can describe the latter as a \emph{two-stage process}. The first stage, purely deterministic, would correspond to the point particle orthogonally ``falling'' onto the membrane, along a rectilinear path, until it reaches its on-membrane position; the second stage, which can either be deterministic or indeterministic (depending on the nature of the membrane), is then the tension-reduction process produced by the disintegration and subsequent contraction of the membrane that we have previously described.

\section{The general tension-reduction (GTR) model}
\label{The general tension-reduction (GTR) model}

In the previous section, we have considered some measurements possibly performed on a coin, generalizing the solipsistic ones usually considered in the classical games of chance. Of course, it was not our intention to describe in a complete and self-consistent way all possible states and measurements that can be described in relation to a coin entity. Our example was just meant to fix ideas and allow introducing some of the basic concepts of a general geometrical description of an entity, which includes not only its states, but also its measurements, and this by means of an interaction mechanism, based on the disintegration of a membrane, which is able to produce the different outcomes and associated probabilities. Based on the intuition we have gained, we are now in a position to reason in more general and abstract terms to identify the fundamental ingredients of what we have called the \emph{general tension-reduction} (GTR) model \citep{AertsSassolideBianchi2015a,AertsSassolideBianchi2015b}.

We begin by summarizing what we have obtained so far, using now a more formal language. Let $\Sigma$ be the set of \emph{all states} of a given entity $S$. By ``all states'' we don't necessarily mean all conceivable states, but more specifically the set of those states that are relevant for the description of the entity in the different measurement contexts it can be meaningfully associated with. If the entity is finite-dimensional, or can be conveniently approximated as such (this is generally the case in all practical experimental situations), which we will assume to be so in the following, then $\Sigma$ can be taken to be a subset of the $M$-dimensional Euclidean space $\real^M$, for some given finite integer $2\leq M <\infty$. 

A measurement on $S$, producing $N$ different outcomes, with $N\leq M$, is then described as a $(N-1)$-dimensional simplex $\triangle_{N-1}$, whose $N$ vertices ${\bf x}_i$, $i=1,\dots, N$, are representative of the $N$ possible outcomes. A measurement of the non-degenerate kind (degenerate measurements will be discussed in Sec.~\ref{Degenerate measurements in the GTR-model}) is then a two-stage process, bringing the initial state ${\bf x}\in\Sigma$ to one of the outcome states ${\bf x}_i$, $i=1,\dots, N$. The first stage of the process corresponds to the point particle associated with the state ${\bf x}$ orthogonally ``falling'' onto the $(N-1)$-dimensional $\rho$-membrane, associated with $\triangle_{N-1}$, and firmly attaching to it. If we write ${\bf x}={\bf x}^\parallel + {\bf x}^\perp$, with ${\bf x}^\parallel$ and ${\bf x}^\perp$ the two components of ${\bf x}$ parallel and orthogonal to $\triangle_{N-1}$, respectively, then this first deterministic stage of the measurement corresponds to the transition: ${\bf x}\to{\bf x}^\parallel$. 

The second stage is a process that can either be deterministic or indeterministic, depending on the nature of the $\rho$-membrane. Its description is a straightforward generalization of what we have explained already in the $N=3$ coin example (see Fig.~\ref{triangolo}). The presence of the point particle on the $\rho$-membrane, at point ${\bf x}^\parallel$, creates $N$ disjoint sub-regions $A_i$, $i=1,\dots N$ ($A_i$ is the convex closure of $\{ {\bf x}_1,\dots,{\bf x}_{i-1},{\bf x}^\parallel,{\bf x}_{i+1},\dots,{\bf x}_N\}$), such that $\triangle_{N-1}=\cup_{i=1}^N A_i$. These sub-regions are separated by $(N-2)$-dimensional ``tension-surfaces,'' along which the elastic substance can less easily disintegrate. Then, as soon as the $\rho$-membrane starts disintegrating, at some point \mbox{\boldmath$\lambda$}, if $\mbox{\boldmath$\lambda$} \in A_i$, the disintegrative process will cause the $N-1$ anchor points ${\bf x}_j$, $j\neq i$, of $A_i$, to tear away, and consequently the $\rho$-membrane will contract toward the remaining anchor point 
${\bf x}_i$, drawing the abstract point particle to that position. Thus, we have a two-stage process producing the possible transitions ${\bf x}\to{\bf x}^\parallel\to{\bf x}_i$, with associated probabilities: 
\begin{equation}
\label{rhoprobabilitymeasurement}
P({\bf x}\to {\bf x}_i|\rho) = \int_{A_i} \rho({\bf{y}})d{\bf{y}}, \quad i\in \{1,\dots,N\}.
\end{equation} 

Let us explore a little further the general structure of $\Sigma$. All points belonging to a measurement simplex $\triangle_{N-1}$ are also possible states of the entity under investigation, as is clear that they are all measurable with respect to a $\rho$-membrane associated with $\triangle_{N-1}$. But it is also clear that within the $(N-1)$-dimensional sub-space generated by $\triangle_{N-1}$, there cannot be other points representative of states in addition to those belonging to $\triangle_{N-1}$ itself. Indeed, \emph{bona fide} states are those that, at least in principle, can participate in all well-defined measurements, but the points in that sub-space lying outside of the simplex cannot be orthogonally projected onto the latter, and therefore cannot be measured with respect to its $\rho$-membrane. 

Of course, this doesn't mean that the points outside of a simplex, in each simplex sub-space, cannot also be used to describe some kind of states, i.e., some real conditions characterizing the entity under study. However, since these states would describe situations where the entity would not be available in producing any outcome (i.e., in providing an answer when subjected to a measurement's interrogative process), they have to be considered states of a non-ordinary kind. Let us call them \emph{confined states}, to express the idea that they describe situations where the entity is confined in a ``place of reality'' which is out of reach for ordinary measurement contexts.

In the example of the coin, we can imagine a situation where the coin has been glued to the wall, with a very strong glue, so that it cannot be subjected anymore to the coin-flipping measurement, or the coin-shaking one. In a cognitive psychology experiment, we can consider the situation of a person who is asked to choose one among a set of predetermined responses to a given question, but with the question and responses expressed in a language that the person cannot understand, so that no meaningful answer can be obtained from her. In physics we can also mention the example of color confinement, the well-known difficulty in directly observing single color-charged entities, like quarks, in our Euclidean spatial theater. Having said that, in the following we will limit our discussion to ordinary (non-confined) states, participating to all possible measurements, which means that by the term ``state'' we will mean (if not mentioned otherwise) a condition in which the entity is available to take part in all well-defined measurements. Consequently, the $(N-1)$-dimensional section of the state space $\Sigma$ that contains the measurement simplex $\triangle_{N-1}$, will be taken to be precisely $\triangle_{N-1}$.

As we said, different measurements are associated with different simplexes, i.e., with simplexes having different relative orientations. This means that the dimension $M$ of the state space $\Sigma$ needs to be large enough to accommodate all these different orientations, in a way that the points belonging to the different simplexes are all mutually orthogonally projectable, so that they can all participate to the different possible measurements. For this to be the case, it is easy to imagine that the dimension $M$ of $\Sigma$ will generally have to be considerably larger than $N$, taking into account the fact that in each subspace generated by a simplex no other simplexes can be present. 

Another issue to be addressed is the center of the simplexes. Of course, they all have to be centered at the same point, say the origin of the system of coordinates considered, otherwise it would not be possible to ensure the overall functioning of the orthogonal projection mechanism. But there is another reason why all measurement simplexes need to share the same origin. The point at the origin corresponds to a state which has quite a remarkable property: it is the state which manifests the same availability in producing whatever outcome, in whatever measurement, and if measurements are described by uniform effective membranes (see Sec.~\ref{The quantum mechanical example}, for the central role played by uniform probability distributions $\rho_u$), it is also the state producing the same probabilities ${1\over N}$, for all outcomes in all measurements. In a sense, it is the most \emph{neutral} state among all possible ones, and if we assume that such condition of \emph{maximum neutrality} should exist, at least in principle, then the different simplexes will have to share the same origin. 

What about the shape of the state space $\Sigma$ in $\real^M$? First of all, what we know is that no states can be at a distance from the origin that is greater than that of the apex points of the different $(N-1)$-dimensional measurement simplexes. Since these points are by definition at distance $1$ from the center of the simplex to which they belong, and that all simplexes share the same center, we have that all the $M$-dimensional vectors of $\Sigma$ are necessarily contained in a $M$-dimensional ball of radius $1$, although of course they will not generally fill such ball. We can also remark that each $(N-1)$-simplex contains an inscribed $(N-1)$-ball of radius ${1\over N-1}$, and since all simplexes have the same origin, within $\Sigma$ there is a $M$-ball of radius ${1\over N-1}$ possessing a maximum density of states, as it contains all the inscribed $(N-1)$-balls associated with the different measurements. Thus, in case the entity under consideration would be associated with a continuity of measurement simplexes, we can expected such $M$-ball to be completely filled with states. Also, considering that for $N=2$ the inscribed 1-sphere (a line segment) has radius $1$, if we have a continuity of two-outcome measurement simplexes $\triangle_{1}$, oriented along all possible directions in $\real^M$, the state space will be precisely a $M$-dimensional ball of radius 1, i.e., $\Sigma = B_1(\real^M)$. 

Of course, apart being contained in a unit ball, and containing a smaller ball having a maximum density of states, nothing can be said a priori about the shape of $\Sigma$, i.e., about the envelope containing the extremal points of $\Sigma$. These extremal points can be of two kinds: either they are at a distance $1$ from the origin, and thus correspond to one of the vertices of a measurement simplex, or they belong to one of the sub-simplexes of a measurement simplex, and then their distance from the origin will be smaller than $1$. This means that, apart the above mentioned two-dimensional case, $\Sigma$ will generally not have a spherical symmetry. 

Strictly speaking, the question of the shape of $\Sigma$ is meaningful only if we have a continuity of states, i.e., if $\Sigma$ is a region of $\real^M$ completely filled with states. In this case, for consistency reasons, it is reasonable to assume that it will be a \emph{convex} region. Indeed, by definition, a convex region is a set of points such that, given any two points, the line joining them lies entirely within it. This means that the region is connected, in the sense that it is possible to go from one point to another without leaving the region. If we assume, as we did, that a measurement is a process during which the state of the entity changes in a continuous way within $\Sigma$, and that both during the first deterministic stage and the second possibly indeterministic stage the abstract point particle representative of the state follows rectilinear trajectories, then the only way to guarantee that in all circumstances these trajectories are made of states, i.e., that they belong to $\Sigma$, is to require $\Sigma$ to be a convex set.

\section{Degenerate measurements in the GTR-model}
\label{Degenerate measurements in the GTR-model}

To complete our description of the GTR-model, we need to consider the possibility of measurements such that different final states can be associated with a same outcome. These are called \emph{degenerate} measurements in quantum mechanics and we will adopt here the same terminology. However, since our approach is more general, we will have to distinguish between two different possibilities, that we will call \emph{submeasurements of the first type} and \emph{submeasurements of the second type} (not to be confused with von Neumann's designation of measurements of the first and second kind). Submeasurements of the first type are degenerate measurements in which the experimenter can in principle distinguish between all the possible outcomes, but decides (for whatever reason) not to do so, thus identifying some of them.

For instance, in the example of the coin, we can imagine the situation where the ``edge'' outcome is conventionally identified with the ``head'' outcome, so that one obtains an effective ``head'' or ``tail'' two-outcome measurement, where ``head'' is now re-interpreted as either point ${\bf x}_{\rm h}$ or point ${\bf x}_{\rm e}$. But apart from this identification, the measurement protocol, and therefore the associated membrane's collapse mechanism, remains exactly the same. In other terms, a submeasurement of the first type, by only identifying some of the outcomes, produces a change of state of the entity that is identical to that produced by an experiment where such identification is not considered. 

Submeasurements of the second type, on the other hand, correspond to experimental situations where the distinction between certain outcomes becomes impossible to realize in practice, even in principle. This means that the outcome states are different than those associated with the corresponding non-degenerate situations, as if they were not, the distinction between the different outcomes would always be possible. In other terms, a submeasurement of the second type is characterized by a different experimental context, and therefore the membrane's mechanism describing its unfolding will also be different. 

In the example of the coin, we can consider the following modified flipping protocol. Once the coin has reached the floor, before taking knowledge of the value of the upper face, a colleague performs the following additional operations: if she finds that the upper face is ``tail,'' she does nothing. If instead she finds that the upper face is ``edge,'' or ``head,'' she takes the coin and places it on a table, tail up, and then without saying a word leaves the room. Clearly, if the final location of the coin is on the floor, the outcome is ``tail,'' and more precisely ``floor-tail''. On the other hand, if the final location of the coin is on the table, then the outcome is ``table-tail,'' and evidently such new state contains no information that would allow the experimenter to associate it either with the ``floor-head'' state or the ``floor-edge'' state, of the associated non-degenerate measurement. 

If we denote $P_{\rm deg}({\rm t})$ the probability of obtaining ``floor-tail,'' and $P_{\rm deg}(\bar{\rm t})$ the probability of obtaining ``table-tail,'' in the degenerate measurement, we clearly have:
\begin{equation}
\label{degenerate-equality}
P_{\rm deg}({\rm t}) = P({\rm t}),\quad P_{\rm deg}(\bar{\rm t})=P({\rm h})+P({\rm e}).
\end{equation} 
Degenerate measurements of the second type which obey equalities of the above kind will be said to be \emph{quantum-like}, as is clear that quantum measurements always obey them. 

Let us now show how we can modify the membrane's mechanism to describe submeasurements of the second type. Starting from the non-degenerate situation, we must alter the functioning of the membrane in such a way that not only (\ref{degenerate-equality}) will be satisfied, but also the collapse will  have to produce the two states ``floor-tail'' (${\bf x}_{\rm t}$) and ``table-tail'' (${\bf x}_{\bar{\rm t}}$), instead of the three states ``floor-tail'' (${\bf x}_{\rm t}$), ``floor-head'' (${\bf x}_{\rm h}$) and ``floor-edge'' (${\bf x}_{\rm e}$). In the following, we only analyze the $N=3$ situation, the generalization to more general measurements being straightforward. 

\begin{figure}[!ht]
\centering
\includegraphics[scale =.55]{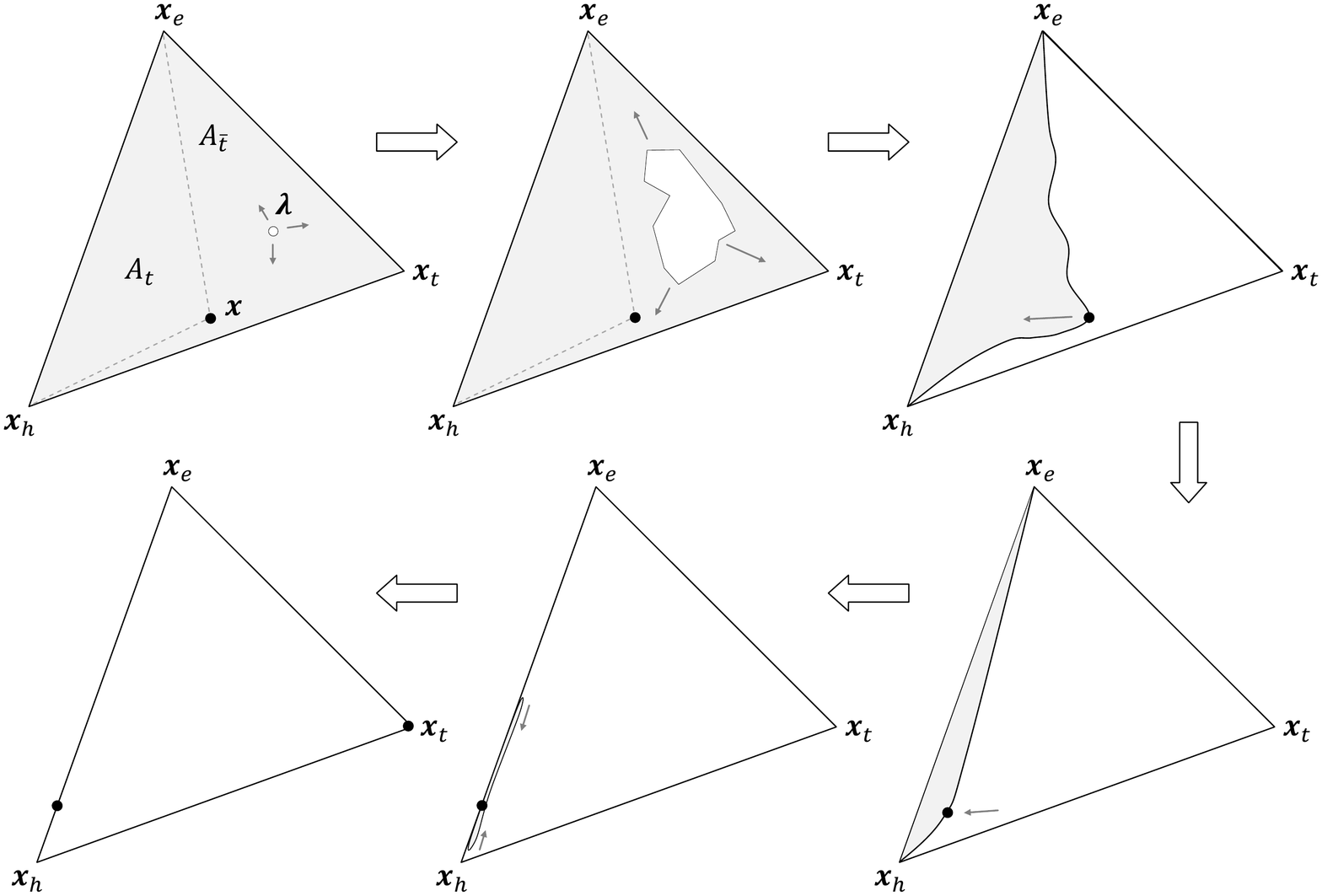}
\caption{The breaking of the elastic membrane (in grey color), when the two regions $A_{\rm h}$ and $A_{\rm e}$ have been fused into a single larger region $A_{\bar{\rm t}}$. The process is here represented in the case where the initial breaking point is in $A_{{\rm h}}$. At first, the membrane disintegrates inside $A_{{\rm h}}\subset A_{\bar{\rm t}}$, causing the anchor point ${\bf{x}}_{\rm t}$ to detach and the particle to be drawn to a position somewhere on the line segment between ${\bf{x}}_{\rm e}$ and ${\bf{x}}_{\rm h}$. Then, also the two anchor points ${\bf{x}}_{\rm e}$ and ${\bf{x}}_{\rm h}$ simultaneously detach, causing the membrane to shrink toward the particle, without affecting its acquired position. Note that the on-membrane state has been denoted ${\bf x}$, instead of ${\bf x}^\parallel$, for notational consistency with Fig.~\ref{triangolo}.
\label{triangolo-degenerate}}
\end{figure}

The first deterministic stage of the measurement is exactly the same as for the corresponding non-degenerate situation, with the point particle initially in state ${\bf x}$ orthogonally ``falling'' onto the membrane and firmly attaching to it. Instead, the second indeterministic stage is different, being that the two sub-regions $A_{\rm h}$ and $A_{\rm e}$ will now be fused together, so as to form a single larger subregion $A_{\bar{\rm t}}=A_{\rm h}\cup A_{\rm e}$. To fix ideas, we can think that a special reactive substance has been applied along the common boundary between $A_{\rm h}$ and $A_{\rm e}$, the effect of this special substance being twofold: firstly, it produces the effective fusion of the two subregions into a single one, so that if the membrane breaks in a point belonging to, say, $A_{\rm h}$, the tearing will now propagate also across the boundary with $A_{\rm e}$ (because of the presence of the reactive substance), causing the collapse of the entire subregion $A_{\bar{\rm t}}$. Secondly, it produces the early detachment of the common anchor point ${\bf{x}}_{\rm t}$, with the consequent contraction of the elastic membrane, drawing the point particle attached to it to a given position on the line segment (the 1-simplex) subtended by ${\bf{x}}_{\rm h}$ and ${\bf{x}}_{\rm e}$. Finally, also the last two anchor points ${\bf{x}}_{\rm h}$ and ${\bf{x}}_{\rm e}$ will detach, causing the membrane to shrink toward the particle, but without affecting its acquired position (see Fig.~\ref{triangolo-degenerate}).

On the other hand, if the membrane breaks in $A_{\rm t}$, then only $A_{\rm t}$ will collapse, producing the final outcome ${\bf x}_{\rm t}$, exactly as in the non-degenerate situation. So, when performing the degenerate measurement, only two transitions are now possibly produced by the membrane mechanism. If the membrane disintegrates in $A_{\rm t}$, the outcome is ${\bf x}_{\rm t}$, which means that the abstract point particle representative of the final state will be at a maximal (unit) distance from the origin of the system of coordinates, that is, from the center of the $M$-dimensional sphere in which $\Sigma$ is inscribed. On the other hand, if the membrane disintegrates in the composite subregion $A_{\bar{\rm t}}$, the membrane's collapse will not necessarily produce a vector of maximal length within $\Sigma$. Therefore, it is natural to assume in this case that a third deterministic process can possibly occur, to complete the measurement, bringing the point particle to a final position ${\bf x}_{\bar{\rm t}}$ of maximal distance from the origin,
at the surface of $\Sigma$, representative of the final outcome-state of the measurement. 

In view of (\ref{rhoprobabilitymeasurement}), it is clear that the degenerate membrane's mechanism we have just described obeys (\ref{degenerate-equality}), considering that the first and third stages of the measurement are purely deterministic. There is however another aspect of interest that needs to be discussed, in relation to this possible third stage of a degenerate measurement of the second type. More precisely, we need to distinguish the following two possibilities. Considering the outcome ${\bf x}_{\bar{\rm t}}$, either (1) its orthogonal projection onto the membrane falls onto some point of the line segment between ${\bf{x}}_{\rm e}$ and ${\bf{x}}_{\rm h}$, or (2) it falls onto whatever other point of the membrane, not lying on that line segment. Condition (1) corresponds to the situation where the corresponding non-degenerate measurement having ${\bf x}_{\bar{\rm t}}$ as the initial state, can only produce the two outcomes ${\bf{x}}_{\rm e}$ and ${\bf{x}}_{\rm h}$, and therefore cannot produce the outcome ${\bf{x}}_{\rm t}$. On the other hand, condition (2) corresponds to the situation where the probability of obtaining ${\bf{x}}_{\rm t}$ will be generally non-zero. 

We shall say that a submeasurement of the second type, obeying condition (1) above, is a \emph{projection-like submeasurement}, as we know that degenerate quantum measurements do obey this condition as a result of the projection formula. A simple way to automatically implement (1) in our model is to ask the third deterministic stage to be structurally similar to the first one, in the sense that the point particle, when reemerging from the membrane, it will always do so following a rectilinear path, orthogonal to the membrane's plane. Such final state, when projected back onto the membrane, in a repetition of the measurement, will then land onto the same edge of the triangular membrane, in accordance with condition (1). It is worth observing, however, that a degenerate measurement of the second type will generally not be projection-like, in the above sense. This is immediately clear in our coin example. Indeed, when flipping the coin in a ``table state'' onto the floor, if we don't use a very special experimental protocol, all three states ``floor-head,'' ``floor-tail'' and ``floor-edge'' will be easily obtained.

\section{Composite entities in the GTR-model}
\label{Composite entities in the GTR-model}

Another important class of measurements we want to describe in our GTR-model is that of measurements performed on \emph{composite entities} (also called \emph{join entities}). To remain within the ambit of our example, consider the situation where instead of a single coin we now have two coins (not necessarily identical). Then, we can perform a first coin-flipping measurement with the first coin, observe the outcome, and do the same with the second coin, observing again the outcome. As each one-entity measurement can produce $3$ different outcomes: ``head,'' ``tail'' and ``edge,'' the combination of the two measurements can produce $9=3\times 3$ different outcomes: ``head-head,'' ``head-tail,'' ``head-edge,'' ``tail-head,'' ``tail-tail,'' ``tail-edge,'' ``edge-head,'' ``edge-tail'' and ``edge-edge.''

It is important to observe that, to be able to perform separately the two measurements, it has to be possible to act separately on the two coins and obtain in a separate way the outcomes of the two experimental actions. We are not saying by this that the two coins must be \emph{experimentally separated}, in the sense that their measurements cannot produce correlations, but that the measurements themselves have to be separable. In physics for instance, when considering measurements on composite entities, these are usually separable, but in more general situations this may not necessarily be the case \citep{AertsSozzo2014a,AertsSozzo2014b}. 

An important characteristic of separable measurements is that they can be performed either sequentially or simultaneously, and that when they are performed in a sequential way the order of the sequence is irrelevant. This is evidently the case in the coin-flipping measurements. Indeed, we can either flip the two coins simultaneously, or one after the other, in whatever order, and we will always obtain the same statistics of outcomes. However, a separable measurement can either produce or not produce correlated outcomes, depending on the nature of the state of the join entity. If such state describes a condition of (experimental) separation of the sub-entities forming the join entity, then no correlations will be observed in the statistics of outcomes. We shall call states of this kind \emph{product states}. The term ``product'' is here to be understood in the specific sense of a state that when subjected to a separable measurement, the obtained set of outcomes can be described as the \emph{Cartesian product} of the set of outcomes obtained when individual measurements are performed separately, on the different sub-entities, in whatever order. 

In the GTR-model, the situation of separable measurements performed on product states can be described by simply considering distinct membranes, one for each sub-entity forming the join entity, working independently from one another. In the case of the two-coin entity, its initial state is then described by a couple of initial states, one for each individual coin, and its measurement is described by a couple of membranes, acting separately on these two one-coin initial states. Of course, it is also possible to represent this double-membrane process as a single higher-dimensional membrane process. For this, instead of two three-dimensional triangular membranes, we will have a 8-dimensional hypermembrane (a 8-simplex), with 9 vertices. The initial state of the two coins will then be described by a vector whose first three components define the state of the first coin, the successive three components define the state of the second coin, and the remaining components (the number of which depends on the dimension $M\geq 8$ of the state space) are fully determined by them (as the join entity state, being a product state, it has to be fully determined by the state of its components). This means that the higher-dimensional single-membrane measurement process, when projected onto its first $3+3$ components, will describe two independent processes, in accordance with the fact that, for a separable measurement performed on a product state, ``the whole has to be equivalent to the sum of its parts.'' 

The situation is however different if either the state is a non-product state, or the measurement itself is a non-separable measurement, or both. For instance, we may decide to change the operational definition of the flipping measurement by introducing a small rigid rod whose two ends are glued to the two coins (in a way that we don't need to specify here), before flipping them. The presence of the connecting rod means that the two-coin measurement cannot anymore be conceived as two separate one-coin measurements, as by flipping one coin we will also, inevitably, flip the other one, and because of the connecting rod (which may break or not during the experiment), correlated outcomes can be observed (which one can show are able to violate Bell's inequalities; see for instance \citet{Aertsetal2000,SassolideBianchi2013b,SassolideBianchi2014}). In other terms, the measurement becomes non-separable. The situation where the state of the two coins is non-product is similar. For instance, one can consider that the two coins are strongly magnetized, and therefore able to attract each other. This means that the description of their state also has to include a description of the magnetic field connecting them, which similarly to the above rigid rod example can produce correlations, under certain conditions. 

Different from the situation of separable measurements performed on product states, in the case of non-separable measurements and/or non-product states, two separate membranes will clearly not be sufficient to obtain a full description not only of the state of the two-coin entity, but also of the statistics of (correlated) outcomes that the measurement is able to produce. Only a genuinely higher dimensional structure will be able in this case to account for all the experimental possibilities, and in particular the $3+3$ components describing the two initial individual coin states will generally not allow to deduce the value of the remaining components, in accordance with the fact that, for non-product (non-separable) entities, ``the whole will be more than just the sum of its parts.''

\section{The quantum mechanical example}
\label{The quantum mechanical example}

In the previous sections we have used the coin example as an heuristic to explain the basic ingredients and structure of the GTR-model. In this section, we consider an important implementation of the GTR-model: \emph{quantum mechanics}. Clearly, the GTR-model is much more general than quantum mechanics, in the sense that it is able to account for a much wider class of measurements and states than those usually considered in the standard quantum formalism. Also, even when the model is reduced to quantum mechanics, by means of two assumptions that we are now going to enunciate, it still remains a more general framework than quantum mechanics, in the sense that it describes a completed version of the latter where the Born rule can be derived in a non-circular way. In that sense, the GTR-model also offers a possible solution to the longstanding measurement problem~\citep{AertsSassolideBianchi2014a}. 

The two additional assumptions that are needed to derive the Born rule of probabilistic assignment from the GTR-model are the following: 

\begin{quotation}
\noindent {\bf Hypothesis 1: Metaignorance}. \emph{The experimenter does not control which specific measurement is actualized at each run of the measurement, among those that can actualize the given outcome-states. In other terms, the experimenter lacks knowledge not only about the (almost deterministic) measurement interaction that each time is actualized, but also about the way it is each time selected.}

\vspace{0.10cm}
\noindent {\bf Hypothesis 2: Hilbertian structure}. \emph{The state space $\Sigma$ is a generalized Bloch sphere.}
\end{quotation}

Let us explain how the above two assumptions can be used to derive the quantum mechanical Born rule. We start by analyzing the consequences of the first one: \emph{metaignorance} (or \emph{metaindifference}, to use the terminology of \citet{Shackel2007}; see also \citet{AertsSassolideBianchi2014b}). This assumption is natural for the following reason. In a typical quantum measurement (like, say, a Stern-Gerlach spin-measurement), we are in a situation where the experimenter doesn't want to control in whatsoever way its development. This because a measurement is generally understood as a process of \emph{observation}, and an observation is meant to alter in the least possible way the observed entity. Of course, if the process creates the very property that is observed, then there will be an ``intrinsic invasiveness'' to it, impossible to remove. This is precisely the situation of quantum measurements, where the entity transitions from an initial to a final state, with the latter being generally different from the former (if the pre-measurement state is not an eigenstate). 

In other terms, in a measurement context the only controlled aspect is that relative to the definition of the possible outcome-states. But apart from that, the experimenter will avoid as much as possible to interfere with the natural process of actualization of these potential outcomes. Thus, underlying the metaignorance assumption there is the idea that if nothing limits the way the measuring system and the measured entity can interact, they will naturally explore all possible (available) ways of interacting. This means that the quantum statistics, and more generally the statistics of any observational process where the experimenter doesn't try to influence the outcomes, corresponds to what has been called a \emph{universal average}~\citep{AertsSassolideBianchi2014a,AertsSassolideBianchi2014b,AertsSassolideBianchi2015a,AertsSassolideBianchi2015b}, i.e., an average over all possible measurement processes associated with a predetermined set of outcomes. Measurements subtending a universal average are called \emph{universal measurements}, and our point is that quantum measurements are just a special example of universal measurements. 

More specifically, to each experimental situation characterized by $N$ given outcomes, an uncountable infinity of measurements can be defined, each one characterized by a different $\rho$-membrane, i.e., by a $(N-1)$-simplex associated with a different probability density $\rho$. A universal measurement then corresponds to a two-stage process, where firstly a $\rho$-membrane is selected among the infinity of possible ones, and secondly, from that specific $\rho$-membrane, a \mbox{\boldmath$\lambda$}-measurement-interaction is also selected, so producing the (almost) deterministic collapse of the membrane. 

So, to calculate the probabilities associated with a universal measurement one has to perform an average over the probabilities obtained from all these different possible measurements. Of course, if such a huge average is addressed directly, one will be confronted with technical problems related to the foundations of mathematics and probability theory. A good strategy, similar to that used in the definition of the \emph{Wiener measure}, is to proceed as follows. First, one shows that any probability density $\rho$ can be described as the limit of a suitably chosen sequence of \emph{cellular probability densities} $\rho_{n}$, as the number of cells $n$ tends to infinity, in the sense that for every initial state ${\bf x}$ and final state ${\bf y}$, one can always find a sequence of cellular $\rho_{n}$ such that the transition probability $P({\bf x}\to {\bf y}|\rho_{n})$, associated with the $\rho_{n}$-membrane, tends to the transition probability $P({\bf x}\to {\bf y}|\rho)$, associated with the $\rho$-membrane, as $n\to\infty$ \citep{AertsSassolideBianchi2014a, AertsSassolideBianchi2015b}. 

By a cellular probability density we mean here a probability density describing a structure made of a finite number $n$ of regular cells (of whatever shape), tessellating the hypersurface of the $(N-1)$-simplex, which can only be of two sorts: uniformly breakable, or uniformly unbreakable. Then, if one excludes the totally unbreakable case (as it would produce no outcomes), we have a total number $2^{n}-1$ of possible $\rho_{n}$-membranes. This means that for each $n$, one can unambiguously define the average probability: 
\begin{equation}
\label{average1}
\langle P({\bf x}\to {\bf y})\rangle_n \equiv {1\over 2^{n}-1} \sum_{\rho_{n}}P({\bf x}\to {\bf y}|\rho_{n}),
\end{equation}
where the sum runs over all the possible $2^{n}-1$ probability densities made of $n$ cells. Clearly, $\langle P({\bf x}\to {\bf y})\rangle_n$ is the probability for the transition ${\bf x}\to {\bf y}$, when a probability density $\rho_{n}$ (a cellular $\rho_{n}$-membrane) is chosen at random, in a uniform way. 

Then, to obtain the transition probabilities of the universal measurement, i.e., of the average over all possible $\rho$-measurements, one has to calculate the infinite cell limit of the above average:
\begin{equation}
\label{average-limit}
\langle P({\bf x}\to {\bf y})\rangle_{\rm{univ}} =\lim_{n\to\infty} \langle P({\bf x}\to {\bf y})\rangle_n,
\end{equation}
and it is possible to demonstrate that \citep{AertsSassolideBianchi2014a, AertsSassolideBianchi2015b}:
\begin{equation}
\label{theoremuniversal}
\langle P({\bf x}\to {\bf y})\rangle_{\rm{univ}} =P({\bf x}\to {\bf y}|\rho_u),
\end{equation}
where $\rho_u$ denotes the uniform probability density, i.e., the probability density associated with a uniform membrane, for which all points have the same probability of disintegrating. 

A model only contemplating uniform membranes has been called the \emph{uniform tension-reduction} (UTR) model \citep{AertsSassolideBianchi2015a,AertsSassolideBianchi2015b}. The result (\ref{theoremuniversal}) then tells us that every time a statistics of outcomes is generated by a universal average, an effective UTR-model will naturally emerge. This is however not sufficient to derive the quantum mechanical Born rule. For this, the second assumption, about the structure of the state space, is also needed. Here again we will not go into any technical details of the proof and just explain in broad terms the gist of it, referring the interested reader to~\citet{AertsSassolideBianchi2014a,AertsSassolideBianchi2015a,AertsSassolideBianchi2015b}.

It is well known that the rays of a two-dimensional Hilbert space can be represented as points at the surface of a 3-dimensional unit sphere, called the \emph{Bloch sphere} \citep{Bloch1946}, with its internal points describing the so-called \emph{density operators} (also called \emph{density matrices}). What is less known is that a similar representation can be worked out for general $N$-dimensional Hilbert spaces \citep{Arvind1997, Kimura2003, Byrd2003, Kimura2005, Bengtsson2006, Bengtsson2013,AertsSassolideBianchi2014a}. The standard 3-d Bloch sphere is then replaced by a generalized $(N^2-1)$-dimensional Bloch sphere, with the only difference that for $N>2$ only a convex portion of it will be filled with states. In other terms, the state space $\Sigma$ is a $M$-dimensional convex set, with $M=N^2-1$, inscribed in a unit sphere of same dimension. 

It is then possible to show that the set of eigenvectors associated with a given observable, i.e., with a self-adjoint operator, are precisely described, within $\Sigma$, by the $N$ vertexes of a $(N-1)$-simplex inscribed in the generalized Bloch sphere \citep{AertsSassolideBianchi2014a}. And when the point particle representative of the initial ray-state, located at some point ${\bf x}$ at a unit distance from the origin, plunges into the sphere to reach its on-membrane position ${\bf x}^\parallel$, following a path orthogonal to the simplex, it can also be shown that this deterministic movement precisely causes the off-diagonal elements of the associated density operator (in the measurement's basis) to gradually vanish, which means that the on-membrane state is precisely a \emph{decohered state}, described by a fully reduced density operator. 

Different from the usual description of \emph{decoherence theory} \citep{Schlosshauer2005}, the decohered on-membrane state ${\bf x}^\parallel$ does not correspond to the final state of the measurement, but to the state prior to the indeterministic collapse of the membrane. In probabilistic terms, the disintegrative/collapse process is governed by the Born rule, as is clear that the coordinates of the on-membrane (reduced density operator) state ${\bf x}^\parallel$ are precisely the transition probabilities and that, as we already remarked in Sec.~\ref{Measuring a coin}, Eq.~(\ref{relative Lebesgue}), the relative Lebesgue measure of the different subregions correspond to the values of the corresponding coordinates of the on-membrane vector. Also, in the situation where the measurement is degenerate, a third purely deterministic process can also happen, of the \emph{purification} kind, through which the entity takes a maximal ``distance'' from the measurement context represented by the membrane, in accordance with the predictions of the L\"uders-von Neumann projection formula \citep{AertsSassolideBianchi2014a}.

To complete our description of the quantum mechanical example, we briefly mention the situation with multipartite systems, formed by multiple entities. For this, we recall that the existence of a generalized Bloch representation is based on the observation that one can always find a basis for the density operators acting in $\compl^N$, made of $N^2$ orthogonal (in the Hilbert-Schmidt sense) operators. One of them is the identity matrix ${\mathbb I}$, and the other $N^2-1$ correspond to a determination of the generators of $SU(N)$, the \emph{special unitary group of degree $N$}. These are traceless and orthogonal self-adjoint matrices $\Lambda_i$, $i=1,\dots, N^2-1$, which can be chosen to be normalized as follows: ${\rm Tr}\, \Lambda_i\Lambda_j=2\delta_{ij}$. Then, any density operator state $D$ can be uniquely determined by a $(N^2-1)$-dimensional real vector ${\bf x}$, by writing \citep{AertsSassolideBianchi2014a}: 
\begin{equation}
D({\bf x}) = {1\over N}\left(\mathbb{I} +c_N\, {\bf x}\cdot\mbox{\boldmath$\Lambda$}\right) = {1\over N}\left(\mathbb{I} + c_N\sum_{i=1}^{N^2-1} x_i \Lambda_i\right),
\label{formulaNxN}
\end{equation}
where we have defined the constant: $c_N\equiv \sqrt{N(N-1)\over 2}$. 

If the entity in question is a joint entity formed by two sub-entities, with Hibert spaces $\compl^{N_A}$ and $\compl^{N_B}$, respectively, so that $\compl^{N}=\compl^{N_A}\otimes \compl^{N_B}$ and $N=N_AN_B$, it is possible to introduce a tensorial determination of the generators, such that the generators of $SU(N)$ are expressed as tensor products of the generators of $SU(N_A)$ and $SU(N_B)$. In this way, one can show that the vector ${\bf x}$ representative of the state of the joint entity can always be written as a direct sum of three vectors \citep{AertsSassolideBianchi2015d}: 
\begin{equation}
{\bf x} = d_{N_A}{\bf x}^{A}\oplus d_{N_B}{\bf x}^{B}\oplus {\bf x}^{\rm corr}.
\label{direct sum}
\end{equation}
where $d_{N_A}=({N_A-1\over N-1})^{1\over 2}$, $d_{N_B}=({N_B-1\over N-1})^{1\over 2}$, ${\bf x}^{A}$ belongs to the one-entity Bloch sphere $B_1(\real^{N_A^2-1})$ and describes the state of the $A$-entity, ${\bf x}^{B}$ belongs to the one-entity Bloch sphere $B_1(\real^{N_B^2-1})$ and describes the state of the $B$-entity, and ${\bf r}^{\rm corr}$ is that component of the state describing the correlations between the two sub-entities.

In accordance with our analysis of Sec.~\ref{Composite entities in the GTR-model}, it is then possible to show that: (1) when the initial state 
${\bf x}$ is representative of a \emph{product state}, the components of the two individual vectors ${\bf x}^{A}$ and ${\bf x}^{B}$ are independent of one another, and the components of the correlation vector ${\bf x}^{\rm corr}$ are entirely determined by the components of the latter. This means that the indeterministic process described by the $(N-1)$-dimensional two-entity measurement's membrane is equivalent to that obtained by two sequential measurements, in whatever order, performed by two $(N_A-1)$- and $(N_B-1)$-dimensional membranes. 

On the other hand, when the initial state is not of the product form $D=D_A\otimes D_B$, the two vectors ${\bf x}^{A}$ and ${\bf x}^{B}$ are not anymore independent and the components of the correlation vector ${\bf x}^{\rm corr}$ cannot anymore be deduced from the components of ${\bf x}^{A}$ and ${\bf x}^{B}$. This means that in a non-product situation the collapse mechanism of the full $(N-1)$-dimensional membrane is needed to describe the statistics of outcomes of the quantum measurement, which cannot be decomposed into two separate and independent collapse mechanisms \citep{AertsSassolideBianchi2015d}.

\section{Beyond Kolmogorov \& Hilbert}
\label{Non-Kolmogorovian non-Hilbertian structures}

In the previous section we have shown that quantum mechanics provides a specific realization of the GTR-model, when the membranes are uniform (a situation we have called the UTR-model) and the state space is a generalized Bloch sphere, which can be seen as the natural completion of the standard Hilbert space structure in which also density matrices are allowed to play the role of pure states. In this section, we provide a few examples of situations where both classical and quantum probabilities structures are violated, showing in this way that the GTR-model can describe more general probability models than those associated with purely classical and purely quantum experimental propositions.

\subsection{Beyond classical}

We start by considering the violation of the classical probability model. More precisely, we will show that the joint probabilities of sequential measurements cannot generally be fitted into a classical probability model equipped with a single sample space. For this, we consider the simplest possible situation: that of an entity whose measurements only have two possible outcomes. The first measurement is characterized by a one-dimensional $\rho_A$-membrane (an elastic band) stretched between the two opposite outcome-states ${\bf a}$ and $-{\bf a}$ (we will call it the $A$-measurement), and a second measurement is characterized by a one-dimensional $\rho_B$-membrane stretched between the two opposite outcome-states ${\bf b}$ and $-{\bf b}$ (we will call it the $B$-measurement). 

We assume that the entity is initially in eigenstate ${\bf b}$ of the $B$-measurement. This means that the probability $P(\to {\bf b}\to {\bf a}|{\bf b})$ of having first the transition to state ${\bf b}$, then to state ${\bf a}$, knowing that the initial state is ${\bf b}$, is: $P(\to {\bf b}\to {\bf a}|{\bf b})=P(\to {\bf b}|{\bf b})P(\to {\bf a}|{\bf b})=P({\bf b}\to {\bf a})$. Also, the probability $P(\to {\bf a}\to {\bf b}|{\bf b})$ of having first the transition to state ${\bf a}$, then to state ${\bf b}$, knowing that the initial state is ${\bf b}$, is: $P(\to {\bf a}\to {\bf b}|{\bf b})=P(\to {\bf a}|{\bf b})P(\to {\bf b}|{\bf a})=P({\bf b}\to {\bf a})P({\bf a}\to {\bf b})$. We thus find that:
\begin{equation}
P(\to {\bf b}\to {\bf a}|{\bf b})-P(\to {\bf a}\to {\bf b}|{\bf b})=P({\bf b}\to {\bf a})[1-P({\bf a}\to {\bf b})].
\label{classicalviolation}
\end{equation}
This means that whenever $P({\bf b}\to {\bf a})\neq 0$ and $P({\bf a}\to {\bf b})\neq 1$, the right hand side of (\ref{classicalviolation}) is different from zero, i.e., $P(\to {\bf b}\to {\bf a}|{\bf b})\neq P(\to {\bf a}\to {\bf b}|{\bf b})$, which is a violation of classical probability, as the (static) propositions of classical probability theory, based on Boolean algebra, always commute. More precisely, in classical theory the probability of the event ``${\bf b}$ then ${\bf a}$'' has to coincide with the probability of the event ``${\bf a}$ then ${\bf b}$,'' i.e., $P_{\rm c}(\to {\bf b}\to {\bf a})= P_{\rm c}(\to {\bf a}\to {\bf b})$, for whatever initial state of the entity under study. Thus, the GTR-model easily violates classical probability. 

\begin{figure}[!ht]
\centering
\includegraphics[scale =.8]{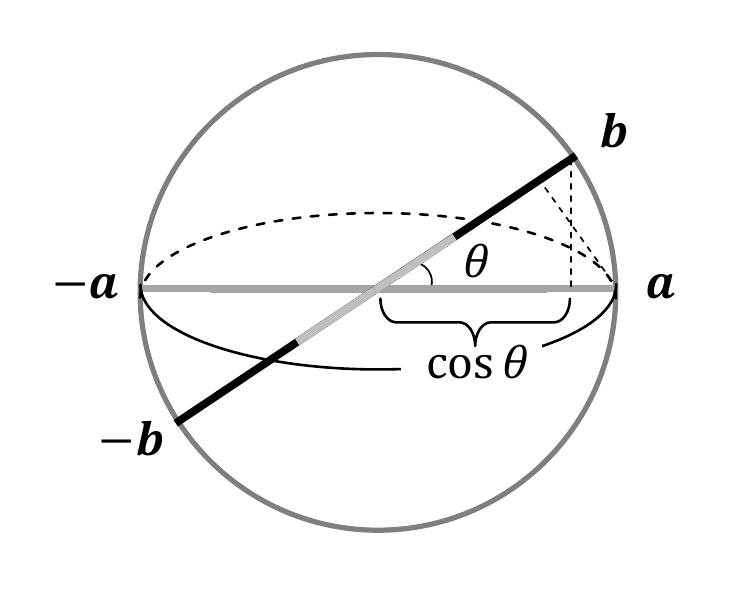}
\caption{Two measurements, characterized by different one-dimensional breakable elastic structures. The measurement having the two outcomes ${\bf a}$ and $-{\bf a}$ is characterized by a uniform $\rho_A$-membrane, whereas the measurement having the two outcomes ${\bf b}$ and $-{\bf b}$ is characterized by a $\rho_B$-membrane that can uniformly break only inside an interval centered at the origin, whose length is strictly less than $2\cos\theta$. If the initial state is ${\bf a}$, all the breakable points of $\rho_B$ will contribute to the transition ${\bf a}\to {\bf b}$. On the other hand, if the initial state is ${\bf b}$, only the points belonging to the segment of length $1+\cos\theta$ will contribute to the transition ${\bf b}\to {\bf a}$.
\label{twoMes}}
\end{figure}

As a specific example, we can assume that $\rho_A$ is uniform, whereas $\rho_B$ describes an elastic band uniformly breakable only inside an interval of length $2\epsilon$, centered at the origin of the sphere, and such that $\epsilon <\cos\theta$, with $\theta$ the angle between the two elastic bands. Then, we have the transition probabilities: $P({\bf a}\to{\bf b})=1$, and $P({\bf b}\to{\bf a})={1\over 2}(1+\cos\theta)$, which are clearly different if $\cos\theta\neq 1$ (see Fig.~\ref{twoMes}).

\subsection{Beyond quantum}
\label{Beyond quantum}

To show that the GTR-model can also easily violate quantum probability, we can still use the example of Fig.~\ref{twoMes}. Indeed, $P({\bf b}\to{\bf a})\neq P({\bf a}\to{\bf b})$ is already a manifest violation of the Born rule, as is clear that according to the latter we should always have the equality (sometimes called the \emph{reciprocity law}): $P({\bf b}\to{\bf a})= P({\bf a}\to{\bf b})$. Indeed, $\langle \psi|\phi\rangle = \langle \phi|\psi\rangle^*$, implying: $P(|\phi\rangle\to|\psi\rangle)=|\langle \psi|\phi\rangle|^2 = |\langle \phi|\psi\rangle|^2=P(|\psi\rangle\to|\phi\rangle)$. This tells us that the transition probability between two states only depends on their relative orientation in the Hilbert space, as measured by the modulus of their scalar product, and not on the specific direction taken by the transition. 

One should not conclude, however, that when the reciprocity law is satisfied the probability model would be Hilbertian. Indeed, as soon as $\rho_A=\rho_B$, we have $P({\bf b}\to{\bf a})= P({\bf a}\to{\bf b})$, but this doesn't mean that the probabilities produced by the elastic bands are necessarily given by the Born rule. For this, as we explained in Sec.~\ref{The quantum mechanical example}, the probability densities have to be uniform. In other terms, testing the reciprocity law is not the same as testing the quantumness of the model, as the reciprocity law can also be satisfied by more general probability models than the Hilbert one. To make this point even more clear, let us introduce a quantity called the \emph{q-test} \citep{BusemeyerBruza2012,WangBusemeyer2013, Wangetal2014}:
\begin{equation}
q \equiv [P(\to {\bf a}\to -{\bf b}|{\bf x}) +P(\to -{\bf a}\to {\bf b}|{\bf x})]-[P(\to {\bf b}\to -{\bf a}|{\bf x})+P(\to -{\bf b}\to {\bf a}|{\bf x})],
\label{q-test}
\end{equation}
where ${\bf x}$ is some given initial state. It can be shown that if the probability model is Hilbertian, then independently of the dimension of the Hilbert space we must have $q=0$, which is usually called the ``QQ-equality'' \citep{BusemeyerBruza2012,WangBusemeyer2013, Wangetal2014, AertsSassolideBianchi2015e}. 

We will give a simple proof of the ``QQ-equality'' in Sec.~\ref{Hilbertian symmetries}. Let us here calculate explicitly the value of $q$ using the GTR-model, to show that the $q=0$ condition can be easily violated. For this, we observe that: $P(\to {\bf a}\to -{\bf b}|{\bf x})= P({\bf x}\to {\bf a})P({\bf a}\to -{\bf b})$, and similarly for the other terms in (\ref{q-test}). Thus, we can write: 
\begin{eqnarray}
\lefteqn{q = [P({\bf x}\to {\bf a})P({\bf a}\to -{\bf b}) +P({\bf x}\to -{\bf a})P(-{\bf a}\to {\bf b})]}\nonumber\\ 
&&\quad - [P({\bf x}\to {\bf b})P({\bf b}\to -{\bf a}) + 
P({\bf x}\to -{\bf b})P(-{\bf b}\to {\bf a})].
\label{q-test2}
\end{eqnarray}
Limiting our discussion to two-outcome situations, we can use $P({\bf x}\to -{\bf b})= 1-P({\bf x}\to {\bf b})$ and $P({\bf x}\to -{\bf a})= 1-P({\bf x}\to {\bf a})$, so that (\ref{q-test2}) becomes: 
\begin{eqnarray}
q&=& q_1+q_2,\\
q_1 &\equiv& P(-{\bf a}\to {\bf b}) - P(-{\bf b}\to {\bf a}) \\ 
q_2 &\equiv& P({\bf x}\to {\bf a})[P({\bf a}\to -{\bf b}) - P(-{\bf a}\to {\bf b})] + P({\bf x}\to {\bf b})[P(-{\bf b}\to {\bf a}) - P({\bf b}\to -{\bf a})].
\label{q-test3}
\end{eqnarray}

The term $q_1$ is called the \emph{relative indeterminism} contribution, and the term $q_2$ is called the \emph{relative asymmetry} contribution \citep{AertsSassolideBianchi2015e}. To simplify the discussion, we assume that all measurements are described by symmetrical probability densities: $\rho_A(y)=\rho_A(-y)$ and $\rho_B(y)=\rho_B(-y)$, as it is the case in the example of Fig.~\ref{twoMes}. Then, we have: $P({\bf a}\to -{\bf b}) = P(-{\bf a}\to {\bf b})$ and $P(-{\bf b}\to {\bf a}) = P({\bf b}\to -{\bf a})$, so that $q_2=0$, but:
\begin{equation}
q_1 =\int_{\cos\theta}^1[\rho_B(y)-\rho_A(y)]dy.
\label{q-test3bis}
\end{equation}
Clearly, if $\rho_B \neq \rho_A$, then $q_1\neq 0$, so that $q\neq 0$, showing again that the probability model described by the GTR-model can extend beyond quantum. 

It is interesting to also observe that being the quantum mechanical situation characterized by uniform probability densities, i.e., $\rho_B = \rho_A = \rho_u$, this means that in a pure quantum model both $q_1$ (relative indeterminism) and $q_2$ (relative asymmetry) are zero. But this is not the only way to satisfy the QQ-equality, as also the condition $q_1=-q_2$ can guarantee that $q=0$. In other terms, the QQ-equality is a necessary but not sufficient condition to test the quantumness of a probability model. 

In fact, even when the stronger condition $q_1=q_2=0$ is satisfied, the model can still be non-Hilbertian. To see this, consider the situation where $\rho_B = \rho_A = \rho$, with $\rho$ a symmetrical (but non-uniform) probability distribution. We then know that, similarly to quantum mechanics, the reciprocal law is satisfied and that $q_1=q_2=0$. As we are now going to show, this doesn't mean however that the probability model is structurally equivalent to that described by the Born rule.

According to the general theorem we have stated in Sec.~\ref{The quantum mechanical example}, only when $\rho$ is a uniform distribution we recover the exact formulae predicted by the Born rule. For example, in the situation of a two-outcome measurement, the Born rule gives: 
\begin{equation}
P_{\rm Born}({\bf a}\to \pm{\bf b})=|\langle \psi_{\bf b}| \psi_{\bf a}\rangle|^2 = {1\over 2}(1\pm\cos\theta),
\label{Born-2-dim}
\end{equation}
which is very different form the GTR-model expression: 
\begin{equation}
P({\bf a}\to \pm{\bf b})= \int_{-1}^{\pm\cos\theta}\rho(y)dy.
\label{GTR-2-dim}
\end{equation}
For instance, for the specific choice $\rho_\epsilon(y)={1\over 2\epsilon}\chi_{[-\epsilon,\epsilon]}(y)$, with $\chi_{[-\epsilon,\epsilon]}$ the characteristic function of the interval $[-\epsilon,\epsilon]$, and $\epsilon\in [0,1]$ (the so-called $\epsilon$-model; see \citet{Aerts1998, SassolideBianchi2013a}), we can write the more explicit expression: 
\begin{equation}
P({\bf a}\to \pm{\bf b}) =\delta_{\pm,-1}\Theta(-\cos\theta -\epsilon) + \delta_{\pm,+1}\Theta(\cos\theta -\epsilon)+\frac{1}{2}\left(1\pm {\cos\theta\over \epsilon}\right)\chi_{[-\epsilon,\epsilon]}(\cos\theta),
\label{probability2b}
\end{equation}
which clearly predicts different values than the Born rule (here $\Theta$ denotes the Heaviside step function, equal to 1 when the argument is positive and equal to 0 otherwise, and $\delta$ denotes the Kronecker delta, equal to 1 when the two indices are the same and equal to 0 otherwise). 

We observe that in the limit $\epsilon\to 0$, $\rho_\epsilon(y) \to \delta(y)$, i.e., the elastic becomes only breakable in its middle point, which corresponds to a measurement with no fluctuations. Then the third term of (\ref{probability2b}) vanishes and one recovers an almost classical situation (almost because for $\cos\theta = \pm\epsilon$ we are in a situation of unstable equilibrium). On the other hand, in the opposite uniform limit $\epsilon \to 1$, $\rho_\epsilon(y)\to {1\over 2}$, the first two terms of (\ref{probability2b}) vanish and the third term tends to the pure quantum expression (\ref{Born-2-dim}).

\begin{figure}[!ht]
\centering
\includegraphics[scale =.75]{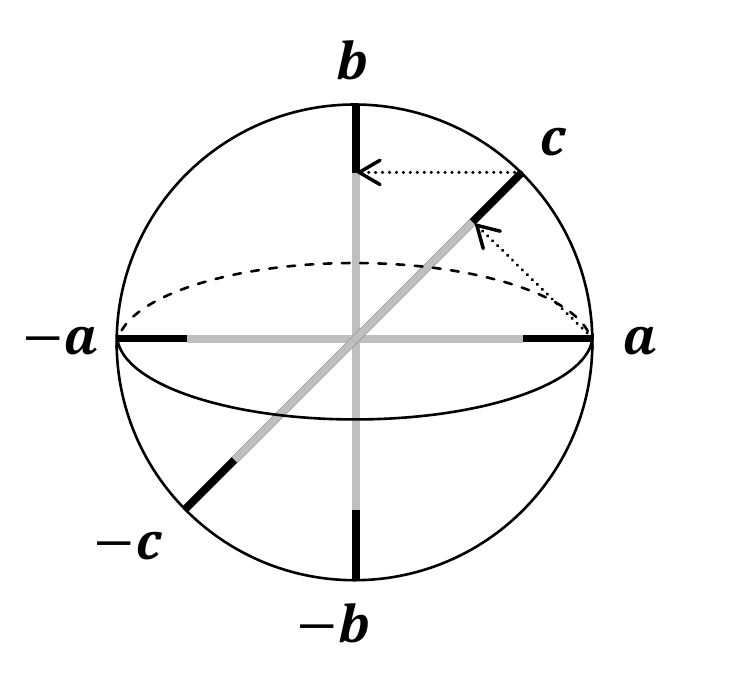}
\caption{Three different measurements represented by three identical symmetric $\rho_\epsilon$-elastic bands oriented along the directions ${\bf a}$, ${\bf b}$ and ${\bf c}$, respectively. The angle between ${\bf a}$ and ${\bf b}$ is $\pi\over 2$, and the angle between ${\bf a}$ and ${\bf c}$, and between ${\bf c}$ and ${\bf b}$, is $\pi\over 4$. The unbreakable segments of the elastics are in black color, the uniformly breakable segments in gray color, and the picture corresponds to the choice $\epsilon ={1\over \sqrt{2}}$ (the breakable segments are therefore of length $\sqrt{2}$).
\label{threeMes}}
\end{figure}

Now, albeit the values of the probabilities predicted by the Born rule (\ref{Born-2-dim}) and by the GTR-model (\ref{GTR-2-dim}) (or more specifically the $\epsilon$-model (\ref{probability2b})) are manifestly different, one may nevertheless ask if the quantum model (i.e., the UTR-model in the Bloch sphere) would nevertheless describe the same experimental situations than a symmetric GTR-model, when all the elastic bands are the same, considering that both models satisfy (at least in the two-outcome situation) the equalities $q_1=q_2=0$. As we are now going to show, the answer is negative. For this, we need to consider three distinct measurements. If they are purely quantum we have, with obvious notation \citep{AertsSassolideBianchi2014a}: 
\begin{equation}
\langle \psi_{\bf a}| \psi_{-\bf b}\rangle = \langle \psi_{\bf a}|\left(| \psi_{\bf c}\rangle \langle \psi_{\bf c}|+| \psi_{-\bf c}\rangle \langle \psi_{-\bf c}|\right)| \psi_{-\bf b}\rangle = \langle \psi_{\bf a}| \psi_{\bf c}\rangle \langle \psi_{\bf c}|\psi_{-\bf b}\rangle + \langle \psi_{\bf a}| \psi_{-\bf c}\rangle \langle \psi_{-\bf c}|\psi_{-\bf b}\rangle,
\label{resolutionidentity}
\end{equation}
where for the first equality we have used the resolution of the identity: $| \psi_{\bf c}\rangle \langle \psi_{\bf c}|+| \psi_{-\bf c}\rangle \langle \psi_{-\bf c}|=\mathbb{I}$. It immediately follows that if we can find an experimental situation in the GTR-model with identical and symmetrical elastic bands such that, say, $P({\bf c}\to -{\bf b})=0$ (implying $\langle \psi_{\bf c}|\psi_{-\bf b}\rangle =0$) and $P({\bf a}\to -{\bf c})=0$ (implying $\langle \psi_{\bf a}| \psi_{-\bf c}\rangle =0$), but also $P({\bf a}\to -{\bf b})\neq 0$ (implying $\langle \psi_{\bf a}| \psi_{-\bf b}\rangle \neq 0$), then such situation would clearly be incompatible with the quantum identity (\ref{resolutionidentity}), and therefore would be modelizable by the symmetric GTR-model, but not by the Born rule of quantum mechanics. A simple situation of this kind is described in Figure~\ref{threeMes}.

\section{The human cognition example}
\label{The human cognition example}

In Sec.~\ref{The quantum mechanical example} we have shown that the quantum mechanical (Hilbertian) model is a special case of the GTR-model, obtained by considering only uniform membranes and a Blochean state space. In this section we want to provide another important implementation of the GTR-model: \emph{human cognition}. In that respect, we recall that as from the beginning of the present century the quantum formalism has been applied with success to model a large number of cognitive phenomena, like information processing, human judgment and decision making, perception and memory, as well as concept combinations and conceptual reasoning; see for instance: \citet{Kitto2008, Khrennikov2010,BusemeyerBruza2012, AertsEtal2013a, AertsEtal2013b, PothosBusemeyer2013, WangEtal2013, BlutnerbeimGraben2014, AertsSassolideBianchi2015a, AertsSassolideBianchi2015b} and the references cited therein.

However, different from the situation of the elementary microscopic entities interacting with macroscopic measuring apparatuses, described by standard quantum mechanics, the general structure and the possible symmetries characterizing the human cognitive activity, when human minds interact with different conceptual entities, remain to be identified. Indeed, even though the quantum formalism has proven to work pretty well as a model, to fit many of the existing experimental data, it also fails to do so for many others (some examples will be given in the following). This means that the general probabilistic structures of the  data generated by the human minds, in the different cognitive contexts, goes beyond that described by pure classical and pure quantum models. Therefore, to study these more general structures one needs an ampler theoretical framework than just classical or quantum mechanics, able to embrace from the beginning the full complexity of these non-physical (mental) processes. The GTR-model provides this more general framework, both in the Chatton and Occam sense: it is not more complex than necessary (Occam's razor), but also it is not less complex than necessary (Chatton's anti-razor). 

Before explaining why the GTR-model provides a natural, coherent and unitary description of human cognition, at a quite fundamental level, it is useful to briefly address a possible objection, which consists in saying that the GTR-model would contain too many free parameters, allowing it to easily fit all sorts of empirical data, but because of this it would not make it of great interest in providing an explanation for the observed phenomena. In fact, considering that an infinity of different probability densities $\rho$ can be used to describe each measurement (in addition to the choice of the orientations of the measurement simplexes), the model clearly allows for an uncountable infinity of free parameters! 

To answer this objection (which by the way is usually also addressed in relation to pure Hilbertian models, obeying the Born rule; see for instance the recent discussion by \citet{BlutnerGraben2015}), one has to distinguish between phenomenological models, where the different parameters are just introduced \emph{ad hoc}, to obtain a good (or exact) data fit, but only for a reduced set of isolated experimental situations, from more fundamental models, designed with the precise intent of describing {\it all possible situations} in a given domain of experimentation, possibly also deriving the observed phenomena from first principles, so providing for them convincing explanations (and whenever possible, predictions). 

To give an example taken from physics, no one would  ever have objected to Einstein that his general relativity theory was a bad explanation because it contained too many free parameters, as for instance the stress-energy tensor appearing in Einstein's field equations could take any functional dependence on the space-time coordinates. Of course, this was not a weak trait of Einstein's model, but its intrinsic richness, considering that his equations had to be applicable to all possible densities and fluxes of energy-momentum in spacetime. And this is precisely what made his theory a universal one. What was important in Einstein's equations is that the free parameters associated with the stress-energy tensor always remained in a clear and logical relation with respect to the other fundamental quantities of the theory, like Einstein's and Ricci's tensors and the metric. And as we are now going to explain, the same holds true, {\it mutatis mutandis}, for the GTR-model. 

When the GTR-model is applied to interrogative contexts involving human minds (a measurement can always be understood as an interrogative context, with the different outcomes being the available answers), different from the standard quantum formalism it allows for a clear distinction not only between `a question and its possible answers,' but also between `the different \emph{ways} an answer can be selected.' Indeed, different human subjects, when subjected to a same interrogation (or situation eliciting a decision), will have each, in general, a different `way of choosing an answer.' Different `ways of choosing' can be described within the GTR-model by means of different probability densities $\rho$ (in the same way as different energy-momentum distributions can be described by means of different stress-energy tensors in general relativity). On the other hand, in standard quantum mechanics all measurements are described by the same uniform probability densities $\rho_u$, which means that a pure quantum model can only describe situations where all subjects participating in the experiment act as perfect ``Bornian clones,'' all selecting an answer exactly in the same way (which is ``the way of a uniform membrane''). 

Also, the disintegration-collapse of a membrane expresses in a very intuitive way what we humans typically perceive when facing a decisional context \citep{AertsSassolideBianchi2015a, AertsSassolideBianchi2015b}. Indeed, when we are subjected to an interrogation, or a situation requiring a decision, we know that at the mental level a (neural) state of equilibrium will be built, expressing a sort of balancing of the tensions between the initial state of the conceptual entity we are subjected to, and the available (mutually excluding and competing) answer-states. The building of this equilibrium is described in the GTR-model by the abstract point particle entering the sphere and reaching an on-membrane position, so producing ``tension lines'' going from its position to the end points representative of the different outcomes. At some moment, always in accordance with what we can subjectively feel, some fluctuations will disturb this mental equilibrium, in a way that we cannot predict in advance, and trigger an irreversible and almost instantaneous process, drawing the abstract point particle to one of the vertices of the mental simplex, reducing in this way the previous tensional equilibrium (hence the ``tension-reduction'' name given to the model, which was suggested to us by Jerome Busemeyer). 

So, the different membranes in the GTR-model are representative of aspects of the minds of the different subjects, understood as dynamic memory structures sensitive to meaning. On the other hand, and consequently, the abstract point particle interacting with a membrane is not to be interpreted (as is usually done) as a description of the subject's beliefs, but as an objective (intersubjective) element of a conceptual reality that is independent of the minds of the individuals that can possibly interact with it. This means that the different locations of the abstract point particle within the generalized Bloch sphere describe the different states a conceptual entity can be in, and that all these states have the same objective status for the different subjects participating in an experiment \citep{AertsGabora2005a,AertsGabora2005b}; but different subjects, because of their different \emph{forma mentis} (their different $\rho$), will extract a different meaning from them, in a given cognitive context, i.e., each subject will choose in a different way an outcome, i.e., an answer to the addressed question (and therefore each subject will be generally associated with a different statistics of outcomes). 

To give an example, consider the concept \emph{Food}. When it is not under the influence of a specific context, we can say that it is in its ``ground'' state, which can be understood as a sort of basic prototype of the concept. But as soon as \emph{Food} is contextualized, for instance in the ambit of the phrase \emph{This food is very juicy}, its state will change. This means that its previous ground state will stop playing the role of a prototype, which will be played then by its new state, in a sort of new `contextualized prototype'. Now, the difference between the concept \emph{Food} in a ground state and in an ``excited'' state, like the one associated with the above ``juicy context,'' can be evaluated by subjecting the concept to an additional context: that of a human mind that is asked to select a good representative of the concept, among a number of possible predetermined choices. The difference between the ground state \emph{Food} and the excited state \emph{This food is very juicy} will then manifest in the fact that, assuming for example that \emph{Fruit} and \emph{Vegetable} are among the possible choices of representatives, the former will be chosen much more frequently (i.e., with a higher probability) than the latter, when the concept is in its ``juicy'' excited state, rather than in its ground state.

\subsection{Replicable measurements}

As we said, a $\rho$-membrane describes in the GTR-model an aspect of a participant's mind subjected to a given interrogative (or decisional) context. It is then natural to consider variations of the probability density $\rho$, when measurements are repeated, to account for the replicability effects that are easy to observe in experimental situations. In quantum mechanics, the replicability of an outcome is only predicted by the theory in relation to the repetition of the same measurement, according to von Neumann first kind condition. As we have seen in Sec.~\ref{Measuring a coin}, the `tension-reduction' mechanism associated with a disintegrable elastic membrane does automatically guarantee that if a measurement is repeated a second time, it will produce exactly the same result, with probability $1$. This because the membrane's collapse cannot alter the position of the point particle when already located in one of the vertices of the measurement simplex. 

However, if, following a measurement $A$, a second different measurement $B$ is performed, and then, following the $B$-measurement, measurement $A$ is performed once again, one will not generally obtain the same outcome obtained in the first $A$-measurement with probability $1$. Indeed, the intermediary $B$-measurement will generally produce an outcome that is not an eigenstate of the $A$-measurement, so that when $A$ is performed a second time the outcome will not be certain in advance and could be different from the first $A$-outcome (unless the two measurements are compatible, a situation described in quantum mechanics by two commuting self-adjoint operators). 

When the $A$ and $B$ measurements are interrogative processes, and the measuring apparatus is a human mind, we know however that the situation is different. Indeed, it is to expected in this case that, in most situations, if we have given a certain answer to question $A$, then we will give the same answer to that same question if we are asked it a second time, even if in the meantime we have also answered to question $B$. This can happen for many reasons, like desire of coherence, learning, fear of being judged when we change opinion, etc. 

In quantum mechanics replicability is easy to model if one assumes that the two measurements $A$ and $B$ are associated with compatible (commuting) observables. The problem is that response replicability is expected to be observed also when the observables $A$ and $B$ are non-compatibles, i.e., when they describe interrogative contexts that, for instance, can give rise to \emph{question order effects}, as commonly observed in social and behavioral research \citep{SudmanBradburn1974, SchumanPresser1981, TourangeauRipsRasinski2000}. In other terms, as \citet{KhrennikovEtal2014} recently emphasized, an experimental situation where both question order effects and response replicability are present cannot be modelized by the standard quantum formalism. 

\begin{figure}[!ht]
\centering
\includegraphics[scale =.38]{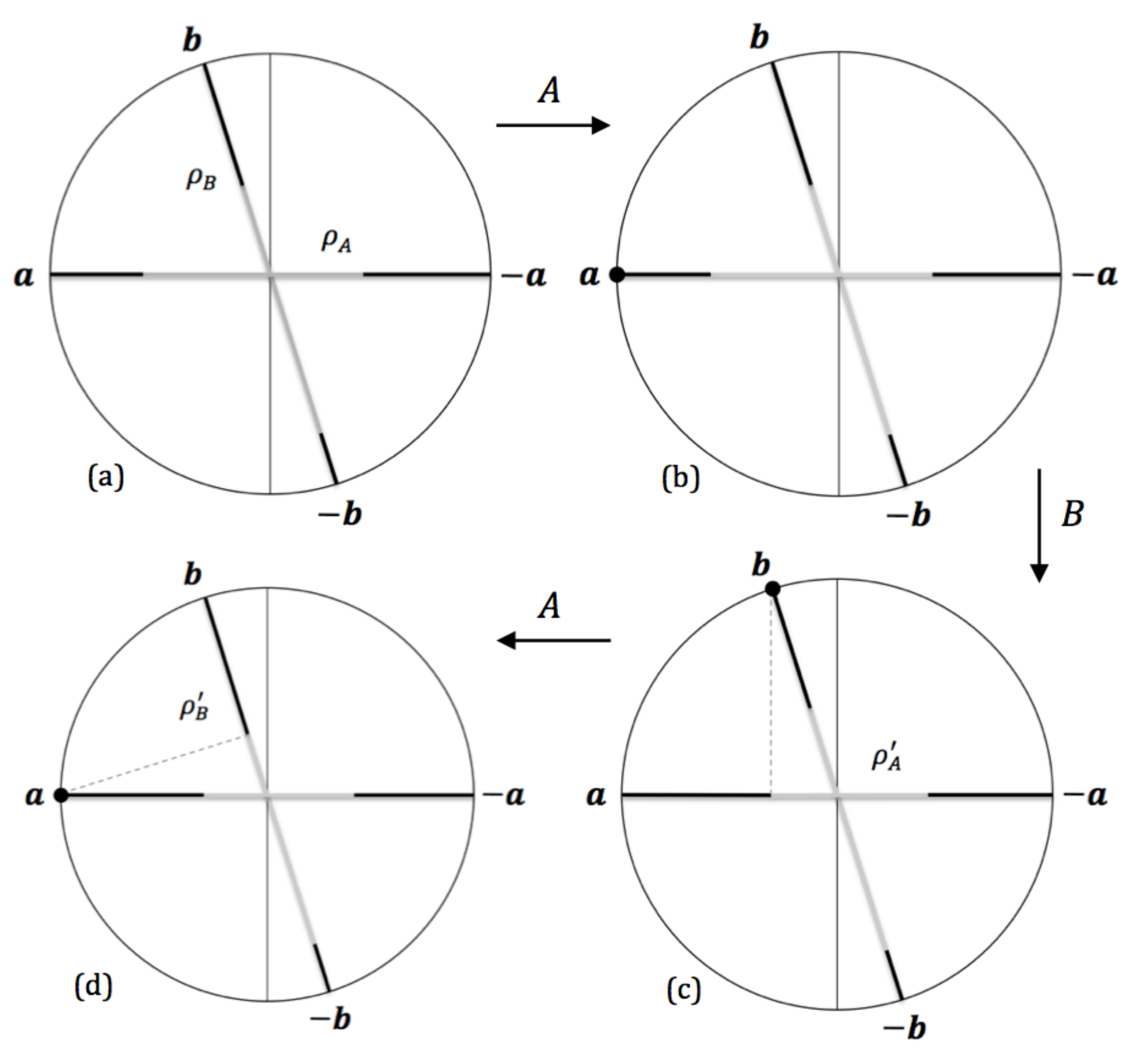}
\caption{A sequence of three two-outcome measurements, $A$, $B$ and then again $A$, in the GTR-model. Figure (a) represents the situation before measurement $A$. The representative point particle is located somewhere at the surface of the sphere, outside of the plane of the two elastic bands describing the two (incompatible) measurements $A$ and $B$, characterized by the non-uniform probability distributions $\rho_A$ and $\rho_B$, respectively. The black parts of the elastics are unbreakable, whereas their grey parts are uniformly breakable. Following the first $A$-measurement, it is here assumed that the indeterministic breaking of the elastic has produced outcome ${\bf a}$, as indicated in Figure (b). Then, in the second $B$-measurement, it is assumed that the transition ${\bf a}\to {\bf b}$ has occurred, as well as the corresponding membrane's change $\rho_A\to \rho'_A$, as illustrated in Figure (c). The outcome ${\bf a}$ of the third $A$-measurement, described in Figure (d), is then certain in advance, in accordance with the hypothesis of replicability, and the ${\bf b}\to {\bf a}$ deterministic transition is also associated with the change $\rho_B\to \rho'_B$ of the $B$-elastic. Subsequent $B$-measurements and/or $A$-measurements will then not change anymore the nature of the elastic bands, and state transitions will become perfectly deterministic.
\label{ABA}}
\end{figure}

What about the GTR-model? Is it able to jointly describe (i.e., jointly model) question order effects and response replicability? The answer is clearly affirmative, as for this it is sufficient to allow the probability density $\rho$, describing the mind aspect of a respondent in relation to a given interrogation, to change in such a way that if the measurement is repeated, following an intermediary measurement, the same answer will be obtained with certainty. In Fig.~\ref{ABA} we give an example of how the probability densities $\rho_A$ and $\rho_B$ characterizing two two-outcome measurements $A$ and $B$ have to change, to guarantee response replicability. 

Note that the elastic bands represented in Fig.~\ref{ABA} are locally uniformly breakable, i.e., uniformly breakable on an interior segment (represented in grey color in the figure) and unbreakable everywhere else (black color in the figure). Note also that the specific structure of these elastics is precisely that required to model in an exact way the data obtained in experiments where subjects were asked to answer the following two incompatible ``yes-no'' questions, producing some typical question order effects \citep{Moore2002,BusemeyerBruza2012, WangBusemeyer2013}: ``Do you generally think Bill Clinton is honest and trustworthy?'' ($A$-measurement) and ``Do you generally think Al Gore is honest and trustworthy?'' ($B$-measurement). We shall not give here the details of this exact modelization, for which we refer the interested reader to \citet{AertsSassolideBianchi2015e}.

\subsection{Non-Hilbertian order effects}
\label{Hilbertian symmetries}

Another example of the insufficiency of the quantum formalism in the modeling of psychological experiments is question order effects. This statement may appear a bit surprising, as these effects are considered by many authors to be among the most successful quantitative predictions of quantum theory in social and behavioral sciences. But this depends on the perspective that is taken on the whole issue. To explain what we mean, we start by considering an equality originally derived by \citet{Niestegge2008} and rediscovered by \citet{WangBusemeyer2013}, which we have discussed already in Sec.~\ref{Beyond quantum}. It is very simple to derive: for this, we denote $P_a$ and $P_b$ two orthogonal projection operators acting on some Hilbert space ${\cal H}$, representing two properties $a$ and $b$ of the entity under study. We also consider the complementary projection operators $P_{\bar a}=\mathbb{I}-P_a$ and $P_{\bar b}=\mathbb{I}-P_b$, describing the orthocomplementary properties ${\bar a}$ and ${\bar b}$, and define the following self-adjoint operator: 
\begin{equation}
Q\equiv P_bP_aP_b - P_aP_bP_a + P_{\bar b}P_{\bar a}P_{\bar b} - P_{\bar a}P_{\bar b}P_{\bar a}.
\label{Q-definition}
\end{equation}
Taking its average $q \equiv{\rm Tr}\, QD$, over an arbitrary state $D$ (which can also be a density matrix), and considering that the probability $P(ba|D)$ of observing, in a sequence, first property $b$ then property $a$, is described in quantum physics by the average \citep{BusemeyerBruza2012}: $P(ba|D)={\rm Tr}\, P_bP_aP_bD$, and similarly for the other sequential (i.e., conditional) probabilities, then using the additivity of the trace, we can write: 
\begin{equation}
q=[P(ba|D) - P(ab|D)]+[P({\bar b}{\bar a}|D) -P({\bar a}{\bar b}|D)],
\label{QQ-equality2}
\end{equation}
which apart from the different notation is precisely (\ref{q-test}). The so-called QQ-equality consists in observing that, for whatever initial state $D$, $q=0$ \citep{BusemeyerBruza2012, WangBusemeyer2013}. This can be easily proven by replacing $P_{\bar a}=\mathbb{I}-P_a$ and $P_{\bar b}=\mathbb{I}-P_b$ into (\ref{Q-definition}), then developing the various terms and see that most of them simplify, so that one is left with the equality $Q=(P_b^2 - P_b) -(P_a^2 - P_a)$. Using the idempotency of the orthogonal projection operators, we thus have $Q=0$, and therefore also $q=0$, for whatever state $D$.

The condition $q=0$ is generally considered to be a good test of the Hilbertian character of the probabilities involved in an experiment. However, as  emphasized in Sec.~\ref{Beyond quantum}, it can also be obeyed  by non-Hilbertian models. This because what we have called the relative indeterminism contribution and the relative asymmetry contribution to $q$, can also mutually compensate, whereas these contributions need to be both zero in a Hilbertian model. But as we have seen, even when these contributions are both zero, the model can still be non-Hilbertian.

This departure from the Hilbertian model can be better appreciated by performing a detailed analysis of data like the Clinton/Gore ones we have mentioned in the previous section. Indeed, although the associated probabilities appear to almost obey the $q=0$ condition, not only there are no reasons to expect that such condition would be exactly obeyed in the ideal limit of an infinite number of participants, but also, and more important, when these probabilities are described using the GTR-model, one immediately sees that their structure is highly non-Hilbertian. This because the breakable elastic bands that are needed to exactly model the data are very different from the globally uniformly breakable structures that are typical of quantum measurements. For example, the elastics depicted in Fig.~\ref{ABA} (a) are precisely those required to generate the Clinton/Gore probabilities, and they are manifestly non-uniform and non-symmetric, i.e., non-Bornian (see also the general discussion in \citet{AertsSassolideBianchi2015e}, where additional quantum identities are derived and shown to be strongly violated by the Clinton/Gore data, and similar ones).

\subsection{Individual and collective minds}

To provide another argument as to why the standard quantum formalism cannot be considered to be sufficient to describe typical psychological measurements, let us come back for a moment to the results of Sec.~\ref{The quantum mechanical example}. We have shown that when different measurements, all having  the same outcomes, are averaged out, in what we have called a universal average, one recovers the Born rule, provided the state space is Hilbertian. A physicists may rightly ask to what exactly these different measurements do correspond. Indeed, when a same measurement is performed a number of times in a physics laboratory, in order to obtain a sufficiently rich statistics of data, and deduce some robust experimental probabilities, the quantum entity is simply prepared every time in the same state, measured by means of the same instrument, in a large number of equivalent runs of the experiment. 

If the interpretation of a quantum measurement as a universal measurements is correct also for physics, this would mean that even though at each run of the measurement the same apparatus is used, the latter would nevertheless each time select an outcome \emph{in a different way}, i.e., according to a different $\rho$-membrane, but since the experimenter would not know which $\rho$ is each time actualized by the apparatus, all these outcomes generated by the different membranes would be averaged out in the final statistics, yielding an effective description in terms of a uniform membrane, which is the Born rule. 

The existence of the hidden-membranes and the hidden-interactions associated with their possible breaking points remains of course hypothetical for the time being in physics, considering that we don't have any direct access to such non-spatial (or  pre-spatial) layer of our physical reality, from our limited Euclidean theater \citep{AertsSassolideBianchi2014a}. This of course does not mean that cleverly designed experiments wouldn't be able in future to reveal these hidden and multidimensional dynamical structures, but for the time being they only remain a compelling theoretical explanation about how the quantum mechanical Born rule can emerge, as a first order approximation, from a substratum of more general probabilistic theories. 

What is the situation in psychological measurements? The main difference is that in that ambit there is an aspect of the measurements that, contrary to quantum measurements in physics laboratories, is not at all hidden. Indeed, in a typical psychological measurement the data are obtained from a number of different participants, for instance about a thousand in the previously mentioned Clinton/Gore experiment. And since these participants are all subjected to the same questions, in relation to a conceptual entity presented to them in the same intersubjective initial state, each participant in the measurement is the manifestation of a different $\rho$-membrane, corresponding to that specific mind aspect characterizing the way each of them, different from all the others, will select one of the available answers. 

In other terms, if the process of actualization of potential $\rho$ remains totally hidden (and also hypothetical for the time being) in physics laboratories, it is instead a perfectly manifest element of reality in psychological measurements. Considering however that the number $n$ of participants is necessarily finite, and in many measurements not  necessarily large, it can be expected that in some situations the obtained average will not be well approximated by a universal one, and therefore the final probability model will not be Hilbertian. To put it in a different way, the abstract ``collective mind'' of the participants may not be representative of a ``pure quantum mind,'' if some ``ways of choosing an outcome'' are not actualizable, because the statistical sample of the available $\rho$-membranes is too small. 

It is however important to emphasize that when we limit our considerations to a single measurement situation, then a Hilbert space probabilistic model, or a Kolmogorovian model, will always be sufficient to fit the experimental data, as these two models are ``universal probabilistic machines,'' capable of representing all possible probabilities appearing in nature, in a \emph{single} measurement context \citep{AertsSozzo2012a,AertsSassolideBianchi2015a}. But when we look for a consistent representation for different non-compatible measurements, this is where classical probabilities become totally inadequate, and pure quantum (Bornian) probabilities become too specific to describe all possible experimental situations.

The situation becomes even more problematic when we try not only to devise a consistent model for the description of a collection of different measurements (different questions) associated with different outcomes (different answers), but when we also consider the possibility of combining these different measurements and their outcomes in a sequential way. This introduces an additional difficulty, which is precisely the problem of distinguishing the individual level from the collective one. Indeed, when psychologists consider sequential measurements, to highlight possible question order effects, it is not the abstract collective mind that is subjected to the sequence of measurements, in different orders, but each one of the individual minds of the participants. In other terms, the sequence of measurements is first performed at the individual level (each participant is asked to answer two questions in a given succession), then an average of their obtained answer is considered. 

Of course, the reason why an overall question order effect is observed is because the effect manifests at the individual level, and is then transferred from the individual to the collective level, in the final statistics of outcomes. If we  ask the first question to an individual, then the second question to another individual,  no order effects would be observed (and the same holds true for response replicability effects). It is of course essential that a same individual in the sample of respondents replies to the two questions in a given order, for the effect to manifest, being it generated at the level of the individual mind and not of the collective one. 

All we are saying is of course perfectly evident, but it is important not to mix these two different levels: the individual and the collective. Let us consider the previously mentioned Clinton/Gore example to further clarify our point. Imagine for a moment that we have found a way to perfectly clone the $i$-th individual participating in the opinion pool. If we repeat many times the ``$A$ then $B$'' and  the ``$B$ then $A$'' sequential measurements, using these $i$-clones (we can only use each clone once, because of response replicability), then calculate the relative frequencies of the observed outcomes and use the GTR-model to fit the data, we would find two elastic bands with a specific orientation in the Bloch sphere, characterized by some generally non-uniform probability densities $\rho_A^{(i)}$ and $\rho_B^{(i)}$. This is the description at the individual level. 

When we consider the responses obtained from all the $n$ different individuals participating in the pool, the probabilities we end up calculating are equivalent to a uniform average over the different sequential probabilities that would have been generated by these $i$-clones, for $i=1,\dots, n$, i.e., the probabilities we can  deduce from the associated $\rho_A^{(i)}$- and $\rho_B^{(i)}$-elastics. These overall probabilities can in turn be modeled by using also two effective elastic bands, characterized by some probability densities $\rho_A$ and $\rho_B$ and a specific orientation. So, at the formal level, the abstract collective mind is described as if it was an individual mind, also performing the sequential measurements, according to its specific ``forma mentis.'' 

However, and this is the subtle point we want to clarify with the present discussion, the collective mind does not really perform a sequential measurement. To see this, consider the following sequential measurement performed ``at the collective level.'' We first ask question $A$ (``Do you generally think Bill Clinton is honest and trustworthy?''), and to obtain an answer we randomly select one of the participants, ask the question and collect the answer. Then, we ask question $B$ (``Do you generally think Al Gore is honest and trustworthy?'') and to obtain the answer we again randomly select one of the participants, ask the question and collect the answer. Similarly, we can perform the same sequential process in reversed order, by asking first $B$ and then $A$. Now, apart the very special circumstance where the same participant would be selected in one or both of the above sequences (a possibility whose probability tends to zero as the number of participants increases), no order effects will be observed in this way. In other terms, at the collective level, if the participants are randomly chosen at each measurement, no order effects will be observed, and of course the same remains true for the response replicability effects. 

This is because if we randomly chose a new participant every time that we ask a question, all memory effects will be destroyed, and all measurements will become compatible. At the individual level, measurements are generally incompatible because the answer given to a first measurement remains in the field of consciousness of the respondent, changing in this way the state of the conceptual entity when a second question is asked. For example, in the Clinton/Gore experiment, the entity which is measured by each individual mind is the conceptual entity \emph{Honesty and trustworthiness} (which for brevity, we shall simply denote \emph{Honesty}). 

Prior to a measurement, we can consider that such entity is in its ``ground'' (most neutral) state, this being true for all the respondents. At the individual level, when performing measurement $A$, a subject is asked if s/he thinks Clinton is honest. Considering this as a measurement of the conceptual entity \emph{Honesty}, the interrogation can be rephrased as follows: ``What best represents \emph{Honesty}, between the two possibilities: \emph{Clinton is honest} and \emph{Clinton is not honest}?'' It is also worth observing that the outcomes  \emph{Clinton is honest} and \emph{Clinton is not honest} are here to be considered as ``excited states'' of the conceptual entity \emph{Honesty}. 

Now, when a subject is submitted to the $A$-measurement, the outcome (i.e., the answer) will generally remain in her/his field of consciousness when the same subject is submitted to the subsequent $B$-measurement, corresponding to the question ``Do you generally think Al Gore is honest?'' This means that when the $B$-measurement is performed immediately after $A$, the measured conceptual entity will not anymore be in its ground state, but in the excited state corresponding to the outcome of the $A$-measurement. If, say, the answer to the $A$-interrogation is ``yes,'' that is, ``Clinton is honest,'' the effective subsequent $B$-measurement will be: ``What best represents \emph{Clinton is honest} between the two possibilities: `\emph{Gore is honest} and \emph{Gore is not honest}?'' 

In other terms, the short term memory of a participant is what allows her/him to keep track of the change of the state of the conceptual entity under consideration, in a sequence of different measurements, and this memory effect, manifesting at the individual level, is what in the end produces the order (and replicability) effects. This memory effect would of course be lost at the collective level, if respondents are selected in a random way at each measurement in a sequence. It could however be restored if the experimental protocol would be so designed to keep track of the obtained answer, and use them as a new input state when a successive measurement is performed, on a new randomly chosen subject. 

Having elucidated this difference between the individual and collective levels, we want now explain how the averaging procedure, when performed on sequential measurements operated at the individual level, can generate a \emph{symmetry breaking process} that can also be held in part responsible for the departure of the experimental probabilities from Hilbertian-like symmetries, like for instance that expressed by the QQ-equality. For this, we consider the simple situation of a collective mind formed by only two individuals and we assume that each of them, when subjected to a sequence of two two-outcome measurements $A$ and $B$, will use the same locally uniform and symmetric probability density, in both measurements. In other terms, we assume that $\rho^{(i)}_A=\rho^{(i)}_B={1\over 2\epsilon_{i}}\chi_{[-\epsilon_{i},\epsilon_{i}]}$, where the index $i=1,2$, denotes the individual subject. Thus, we are here in a situation where both the relative indeterminism contribution and the relative asymmetry contribution are zero and the symmetry expressed by the QQ-equality is obeyed. In other terms, even though the measurements are not purely quantum, they still obey the QQ-equality that is also obeyed by pure quantum systems, and in that sense (but only in that sense) the situation can be considered to be, at the individual level, close to a pure quantum situation. 

What happens then when we consider a uniform average over these two participants? To see this, we denote ${\bf a}$ and $-{\bf a}$ the two outcomes of measurement $A$, and ${\bf b}$ and $-{\bf b}$ those of measurement $B$, as represented in the Bloch sphere. According to (\ref{probability2b}), if we assume that the angle $\theta$ between the two elastic bands ($\cos\theta = {\bf a}\cdot {\bf b}$) is such that $\cos\theta <\epsilon_{i}$, $i=1,2$, and that the angle $\theta_A$ between the unit vector ${\bf x}$ describing the initial state and outcome ${\bf a}$ ($\cos\theta_A = {\bf x}\cdot {\bf a}$) is such that $\cos\theta_A <\epsilon_{i}$, $i=1,2$, then for the sequence of outcomes ``${\bf a}$ then ${\bf b}$'' we have the $i$-individual probability: 
\begin{eqnarray}
P^{(i)}(\to {\bf a}\to -{\bf b}|{\bf x})&=& P^{(i)}({\bf x}\to {\bf a}) P^{(i)}({\bf a}\to {\bf b})= {1\over 2}(1+ {1\over \epsilon_i}\cos\theta_A){1\over 2}(1+ {1\over \epsilon_i}\cos\theta)\nonumber\\
&=&{1\over 4}[1+{1\over \epsilon_i}(\cos\theta +\cos\theta_A)+{1\over \epsilon_i^2}\cos\theta\cos\theta_A].
\label{symmetricalprob}
\end{eqnarray}
If now we consider the uniform average: $P(\to {\bf a}\to -{\bf b}|{\bf x})\equiv{1\over 2}[P^{(1)}(\to {\bf a}\to -{\bf b}|{\bf x})+P^{(2)}(\to {\bf a}\to -{\bf b}|{\bf x})]$, we obtain: 
\begin{equation}
P(\to {\bf a}\to -{\bf b}|{\bf x})= {1\over 4}\left[1+{\epsilon_1 +\epsilon_2\over 2\epsilon_1 \epsilon_2}(\cos\theta +\cos\theta_A)+{\epsilon_1^2+\epsilon_2^2 \over 2\epsilon_1^2\epsilon_2^2}\cos\theta\cos\theta_A \right].
\label{average-n=2}
\end{equation}

We  observe that (\ref{average-n=2}) can be written in the form (\ref{symmetricalprob}) only if $(\epsilon_1 +\epsilon_2)^2=2(\epsilon_1^2+\epsilon_2^2)$, or $(\epsilon_1-\epsilon_2)^2 =0$, i.e., if $\epsilon_1=\epsilon_2$. But since by hypothesis $\epsilon_1\neq\epsilon_2$, we immediately see that the effective elastic bands describing the two measurements from the viewpoint of the ``collective mind'' (here just formed by two individuals), not only they are not anymore the same, but also they are not anymore symmetric (for a specific calculation, see \citet{AertsSassolideBianchi2015e}). Also, the obtained averaged probabilities will generally violate the QQ-quality. In other terms, the averaging procedure induces a breaking of possible symmetries in the structure of the probabilities, and a clear departure from the Hilbertian model.

A remark is in order. In Sec.~\ref{The quantum mechanical example}, we have explained that when all possible membranes (i.e., all possible ways of choosing an outcome) are allowed to be actualized in a single (non-sequential) measurement context (in what we have called a universal average, or universal measurement),  the averaged probabilities are then described by an effective uniform membrane, equivalent to the Born rule, when the state space is considered to be Hilbertian. As we have just seen, the process of averaging over different membranes becomes much more involved when we deal with sequential measurements. Certainly, when considering a finite number of participants (the actual situation in real experiments), not all behaving as ``Bornian clones,'' the final statistics will be non-Hilbertian. However, it remains an open question to determine what would be the probability model of a `universal sequential measurement,' i.e., of an average over sequential measurements when all possible probability densities are included in the calculation. We plan to come back to this interesting question in future works.

\section{Concluding remarks}
\label{Conclusion}

In the present work we have reviewed and further illustrated some of the results we have recently obtained in \citet{AertsSassolideBianchi2014a, AertsSassolideBianchi2014b, AertsSassolideBianchi2015a, AertsSassolideBianchi2015b, AertsSassolideBianchi2015c, AertsSassolideBianchi2015d, AertsSassolideBianchi2015e, AertsSassolideBianchi2015f}, to emphasize the interest and role played by the GTR-model (and the associated hidden-measurement approach) in the description of very general measurement situations, extending beyond the pure classical and pure quantum ones. As we have explained, these more general situations, and the associated probability models, are certainly relevant in the description of both physical and psychological experimental situations. 

Classical (Kolmogorovian) probabilities generally describe ``static propositions,'' i.e., our lack of knowledge about the actual elements of reality that are present in the system under study. On the other hand, quantum probabilities generally describe ``dynamic propositions,'' i.e., our lack of knowledge about processes of actualization of potential properties (if we exclude the special situations of measurements performed on eigenstates). Somehow in between these two descriptions, we can consider mixed measurements, where both static and dynamic logics, discovery and creation processes, actuality and potentiality, determinism and indeterminism, play an equivalent role. These more general, hybrid contexts, cannot be described using the too limited Hilbertian or Kolmogorovian models, but require more general structures.

In that respect, it is important to realize that our reality, because of its extreme complexity, is able to manifest all sorts of mixtures of creation and discovery processes, also at a fundamental level. Therefore, in our investigations we need to be equipped with probabilistic models that are able to cope with such complexity, beyond the very specific classical and quantum structures. This of course does not mean that, in certain ambits, one will not try to highlight some possible remarkable symmetries, but to correctly describe them we certainly need a general enough theoretical approach, as only in this way we can hope to understand the full logic behind them. The example of the QQ-equality is in that sense paradigmatic: we know that the equality is exactly obeyed by pure Hilbertian models, but we have seen it can also be obeyed by a class of non-Hilbertian models, which are precisely those describing the question order effect in ``opinion pool'' psychological measurements. This means that if we want to understand the reasons behind these observed regularities, we cannot do so from the limited Hilbertian viewpoint, but need a more general approach. One of the scope of the present article was to point out that the GTR-model precisely provides such needed more general approach. 

Actually, we think that the GTR-model does more than this. Indeed, if it is correct to say that the Kolmogorovian model is a universal model for the description of situations governed by ``static information,'' certainly we cannot say that the Hilbertian model, equipped with the Born rule, is a universal model for situations governed by (non-Boolean) ``dynamic information,'' where also lack of knowledge about processes of actualization of potential properties is considered. This not only because the state space is Hilbertian, which may be a too severe constraint in certain situations, but also because very specific collapsing membranes are considered in quantum mechanics: the uniform ones. When all possible structures of membranes and state spaces are allowed, one certainly obtains the most general possible probabilistic description, i.e., a universal model for the description of both static and dynamic situations, which we have called the  GTR-model.

As our simple coin example illustrates, the necessity of using the more general GTR-model already manifests when considering experimental situations involving macroscopic objects. This because classical properties, and the associated classical probabilities, are insufficient to describe all possible observations. In fact, macroscopic objects possess more physical properties than those usually accounted for by classical mechanics, like for instance the ``upper face'' property of a coin, or of a die \citep{SassolideBianchi2013b}. And when these non-ordinary properties are considered, and tested in an operational way, not only the classical Kolmogorovian probabilities become inadequate, but the Hilbertian (Bornian) ones as well. On the other hand, the universality of the GTR-model allows to properly handle these non-ordinary measurement situations, and we cannot exclude that its structural richness will not also be instrumental in the description of anomalies manifesting in measurements with elementary physical entities (see for instance \citet{AdenierKhrennikov2007} for an example of possible anomalies in the ambit of coincidence measurements). 

However, if the interest of the GTR-model for elementary (microscopic) physical systems remains to be evaluated (apart of course its theoretical interest in deriving and explaining the Born rule as a universal average), it is already a necessary tool for properly modeling human cognition, as we have illustrated in this work, considering that the data already in our possession cannot be exactly fitted by means of classical and Hilbertian models. We therefore hope that more scientists will decide to adopt it with advantage, both conceptually and as a mathematical instrument, to explore the ubiquitous quantum-like (but not necessarily pure quantum) structures.

\end{document}